\documentstyle[11pt,psfig]{article}

\makeatletter
\if@twoside
\else \oddsidemargin 00pt \evensidemargin 00pt
\fi
\def\section{\@startsection {section}{1}{\z@}{+6.0ex plus +1ex minus
 +.2ex}{2.8ex plus .2ex}{\large\bf}}
\def\subsection{\@startsection {subsection}{2}{\z@}{+3.0ex plus +1ex
minus +.2ex}{2.3ex plus .2ex}{\normalsize\bf}}
\def\subsubsection{\@startsection{subsubsection}{3}{\z@}{+2.5ex plus
+1ex minus +.2ex}{1.5ex plus .2ex}{\normalsize\bf}}
 
\def\theequation{\thesection.\arabic{equation}}
\@addtoreset{equation}{section}

\@addtoreset{figure}{section}

\@addtoreset{table}{section}
 
\def\appendix{\par
 \setcounter{section}{0} \setcounter{subsection}{0} 
 \def\thesection{\Alph{section}}}
 
\expandafter\ifx\csname mathrm\endcsname\relax\def\mathrm#1{{\rm #1}}\fi
\@ifundefined{mathrm}{\def\mathrm#1{{\rm #1}}}{\relax}
\makeatother
 
\newcommand{\SLASH}[2]{\makebox[#2ex][l]{$#1$}/}

\newcommand{\kslash}{\SLASH{k}{.15}}

\newcommand{\id}{{\rm 1\kern-.12em
\rule{0.3pt}{1.5ex}\raisebox{0.0ex}{\rule{0.1em}{0.3pt}}}}
\newcommand{\D}{\displaystyle}
\newcommand{\T}{\textstyle}
\renewcommand{\Re}{{\rm Re\,}}

\newcommand{\st}{{s_\theta}}
\newcommand{\sd}{{s_\delta}}
\newcommand{\ct}{{c_\theta}}
\newcommand{\cd}{{c_\delta}}
\newcommand{\Hn}{{H^0}}
\newcommand{\Hpm}{{H^\pm}}
\newcommand{\Kn}{{K^0}}

\newcommand{\beq}{\begin{equation}}
\newcommand{\eeq}{\end{equation}}
\newcommand{\bea}{\begin{eqnarray}}
\newcommand{\eea}{\end{eqnarray}}
\newcommand{\G}{\Gamma}
\newcommand{\al}{\alpha}

\newcommand{\nn}{\nonumber}
\newcommand{\mz}{M_Z^2}

\pssilent

\begin{document}
\thispagestyle{empty}
\hfill KA-TP-3-1997

\hfill hep-ph/9703392

\vspace{2cm}
\begin{center}

{\LARGE \bf
Precision observables in SU(2)$\times$U(1) models \\
with an additional Higgs triplet}

\vspace{2cm}

{\large \sc T. Blank and W. Hollik}

\vspace{2cm}

{\sl Institut f{\"u}r Theoretische Physik}

{\sl Universit{\"a}t Karlsruhe}

{\sl Kaiserstra\ss{}e 12}

{\sl 76128 Karlsruhe, Germany}

\end{center}

\vspace{1cm}

\begin{abstract}
\noindent Electroweak precision observables
are calculated at complete 1-loop order 
 in the extension of the standard model by an
extra Higgs triplet, where the $\rho$-parameter can be different
from unity already at the tree level. 
One additional data point is required for fixing the input
parameters. In the on-shell renormalization scheme
the leptonic mixing angle $\sin^2\theta_e$ at the $Z$ peak is chosen, 
together with 
the conventional input $\alpha, M_Z, G_{\mu}, m_t$. 
The calculated observables depend on the 
mass of the doublet Higgs boson $H^0$ and on the masses of the extra 
non-standard Higgs bosons as free parameters.
The predictions of the 
standard model and the triplet model
coincide for all observables in the
experimental range of the top mass $m_t=175\pm 6$ GeV.
In the triplet model,  all observables which show
a dependence on the doublet Higgs mass $M_{H^0}$ are consistent with a low
value of $M_{H^0}$.
\end{abstract}

\newpage

\thispagestyle{empty}
\clearpage
\setcounter{page}{1}

\section{Introduction}
\noindent 
In the light of the recent electroweak precision data
the standard model with a single Higgs doublet is in a very good shape
\cite{Blondel:96}.
Whereas the data are compatible with a relatively light Higgs boson,
direct empirical information on the scalar sector, however, 
is still lacking. A specific feature of the
standard model assumption of a single Higgs field is the validity of the
tree level relation
$$ 
   \rho \ = \ \frac{M_W^2}{M_Z^2\ \cos^2\theta_W} \ = \ 1 
$$
for the $\rho$-parameter, which measures the ratio between the 
neutral and charged current coupling strength \cite{ross}.
$\rho$ deviates from unity by electroweak quantum effects, 
especially from the top-bottom doublet \cite{veltman}.
In more general scenarios, there are already tree level contributions
to $\rho-1$, 
 which however can only be of the order of the standard loop effects
not to spoil the
agreement with experimental data. 
A consistent formulation of such a scenario with 
$\rho_{tree} \neq 1$ requires the extension of the Higgs sector
by at least an additional triplet of scalar fields with one extra
vacuum expectation value different from zero. 
The full set of precision observables can be calculated,  
in analogy to the minimal model, 
 in terms of a few
input data points together with the standard loop contributions and the
loops arising from the non-standard  Higgs part.
A complete discussion of the radiative corrections requires not only 
the evaluation of the extra loop diagrams with non-standard 
Higgs bosons, but also an extension of the renormalization
procedure. Since $M_W,M_Z$ and $\sin^2\theta_W$ are now independent
parameters, one extra renormalization condition is necessary.
This can be chosen in a formal way as done in the 
$\overline{MS}$-scheme \cite{passarino}, or in extension of the
standard on-shell scheme \cite{hollik89}
by choosing the electroweak mixing angle at the 
$Z$ peak, $\sin^2\theta_e$ for leptons, as an additional input 
parameter, together with the usual input $\alpha, G_F, M_Z$.

\smallskip
In this paper we give  a complete one-loop calculation of
 $M_W$ and the $Z$ boson observables in the simplest extension
of the minimal model accommodating $\rho_{tree} \neq 1$.
This model 
(discussed to some extent also in \cite{Lynn:92}) 
augments the standard model 
by an additional Higgs triplet with a
 VEV $\neq 0$ in the neutral sector. Besides the standard Higgs boson $H^0$
a further neutral
scalar boson $K^0$ and a pair of charged Higgs particles $H^{\pm}$ 
form the physical spectrum.
After specifying the model in section 2, we outline in section 3 the 
calculation in the aforementioned extended on-shell scheme.
The predictions for the various observables and their
parameter dependence are discussed and compared with the standard model
predictions as well as with the experimental data in sections 4 and 5.
Details of the calculation are collected in the appendix.

\section{The standard model with an extra Higgs triplet}
\label{model}
\noindent
We consider the 
extension of the electroweak standard model where  
besides the ordinary Higgs doublet field 
\begin{equation}
  \Phi(x) = \left( 
  \begin{array}{c}
    \phi^+ (x)\\ 
    \frac{1}{\sqrt{2}}(v+H^0(x)+i\chi (x))
  \end{array}\right) 
\label{Higgsdublett}
\end{equation}
an additional Higgs field $\Delta$ is  introduced
which transforms as a triplet under the symmetry group 
SU(2)$\times$U(1). Couplings of this extra field to fermions, although
possible  \cite{Gunion:90}, are not  considered for simplicity.
The hypercharge is assigned as $Y_\Delta = 0$, thus no particles with
double electric charge occur. 
With a vacuum expectation value $u$ in the
neutral component, the triplet can be written as \cite{Lynn:92}
\begin{equation}
  \Delta = \left(
  \begin{array}{c}
    \Delta^+ \\ \Delta^0 = u + K^0 \\ \Delta^-
  \end{array}\right) \;
  \quad \mbox{with} \quad {\Delta^0}^* = \Delta^0\,,\,
  {\Delta^+}^* = - \Delta^- \,.
  \label{Higgstriplett}
\end{equation}
Since there is no need for Higgs self couplings in our calculations,
we can restrict our discussion to the extra Higgs term in the kinetic
part of the Lagrangian
\begin{equation}
  {\cal L}_{\Delta-kin} = \frac{1}{2} (D_\mu\Delta)^\dagger(D^\mu\Delta) .
\end{equation}

The unphysical Higgs fields $G^{\pm}, G_Z$ and the charged physical 
Higgs $H^{\pm}$ are linear combinations of the doublet and triplet
field components
\begin{equation}
  \left(
    \begin{array}{c}
      G^\pm \\ H^\pm
    \end{array}
  \right) =
  \left(
    \begin{array}{cc}
      \cos{\delta} & \pm\sin{\delta} \\
      -\sin{\delta} & \pm\cos{\delta}
    \end{array}
  \right)
  \left(
    \begin{array}{c}
      \phi^\pm \\ \Delta^\pm
    \end{array}
  \right) \quad,\quad G_Z=\chi\quad,
\label{Higgs_Rotation}
\end{equation}
where the mixing angle $\delta$ is determined by the vacuum
expectation values $u$ and $v$:
\begin{equation}
  \cos^2\!\delta = \frac{v^2}{v^2+4 u^2} \, .
\end{equation}
Besides the standard Higgs $H^0$, there is a further neutral physical
Higgs field $K^0$.
In the Feynman-'t Hooft gauge the unphysical fields $G^\pm$ and
$G_Z$ get the same masses as the corresponding vector bosons. The
masses of the remaining physical fields $H^0, K^0, H^\pm$ 
are free parameters.

In this model, in the following denoted as triplet model (TM), 
the
masses of the $Z$ boson and the photon follow from $v$ as in the SM
\begin{equation}
  M_A = 0  \quad,\quad M_Z = \frac{1}{2} \sqrt{g_1^2+g_2^2}\; v
  \quad,
\end{equation}
but due to the additional vacuum expectation value $u$, the $W$-mass
has changed to
\begin{equation}
  M_W = \frac{1}{2}\, \frac{g_2 v}{\cos\delta} \quad.
  \label{TM_VB_Massen}
\end{equation}
The electroweak mixing angle, which diagonalizes the neutral gauge boson
mass matrix, is determined by  
\begin{equation}
  \cos{\theta_W^{Triplett}} =
  \frac{g_2}{\sqrt{g_1^2+g_2^2}}
     =: \ct  \quad,\quad \st^2 = 1 - \ct^2 \, .
\label{TM_Mischungswinkel}
\end{equation}
It is related to the quantity (the mixing angle in the minimal model) 
\begin{equation}
  c_W =\frac{M_W}{M_Z} \quad,\quad s_W^2=1-c_W^2
  \label{SM_Weinberg}
\end{equation}
in the following way:
\begin{equation}
  c_\theta = c_W \cos\delta \quad.
\label{Definition_c_theta}
\end{equation}
This means that for $u\neq 0$,
the $\rho$-parameter is 
different from unity already at the tree level: 
\begin{equation}
  \rho\, \equiv \, \frac{M_W^2}{M_Z^2 c_{\theta}^2} =
  \frac{1}{\cos^2\delta}\quad.
  \label{rho-cdelta}
\end{equation}

\section{One-loop calculations and renormalization}
\label{renormalization}
\noindent 
In order to obtain finite amplitudes in the TM at the 1-loop level we 
perform the renormalization in an on-shell scheme which is similar to
the one described in \cite{hollik89} for the minimal SM. 
Compared to the minimal model, the TM has one more independent parameter
in the gauge boson - fermion sector, which may be chosen as $\rho$ or
$\st^2$. For the renormalization procedure it is more convenient to
treat $\st^2$ as an
additional independent input parameter and fix its counter term 
$\delta\st^2$ by an appropriate renormalization condition.
The other basic on-shell parameters with independent counter terms are
$M_W, M_Z$ and the electric charge $e$, which are renormalized by the
same set of conditions as in the minimal model \cite{hollik89}.
$\rho$ then appears as a derived quantity.

The renormalized vector boson self energies at the 1-loop level have the
following counter term structure:\hfill\\[1cm]
\begin{tabular}{p{3.3cm}@{\hspace{0.5cm}=\hspace{0.5cm}}p{3.3cm}
    @{\hspace{0.5cm}+\hspace{0.5cm}}p{6cm}}
  \centerline{\raisebox{-0.48cm}{\psfig{bbllx=131pt,bblly=337pt,bburx=481pt,bbury=455pt,figure=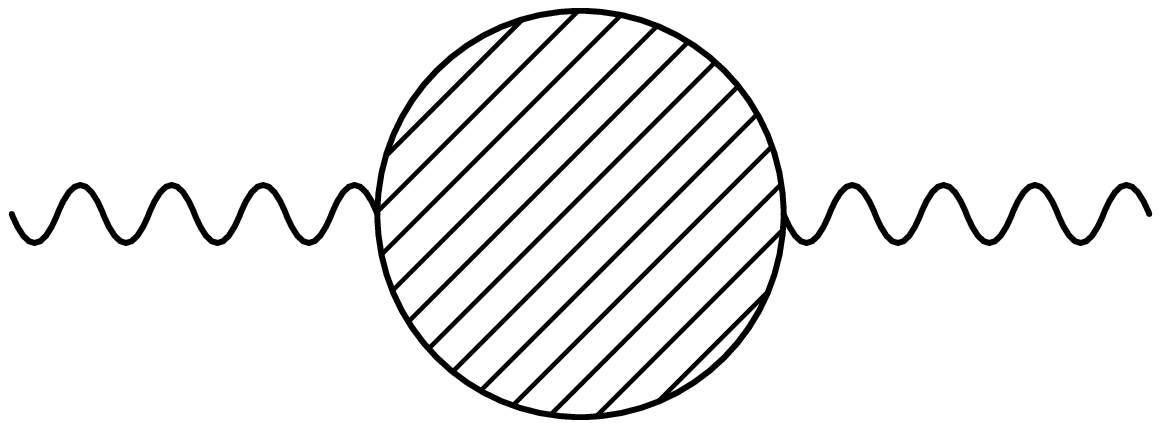,width=3cm}}} &
  \centerline{\raisebox{-0.48cm}{\psfig{bbllx=131pt,bblly=337pt,bburx=481pt,bbury=455pt,figure=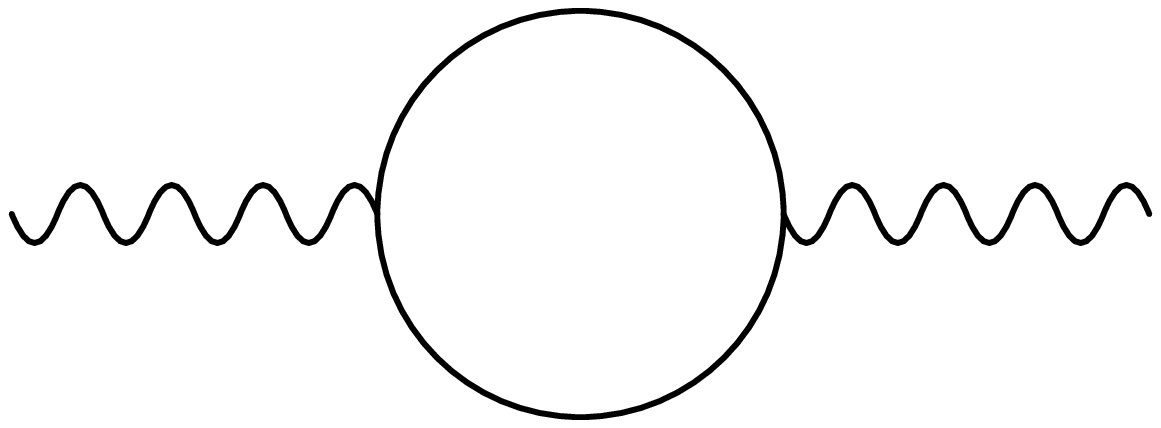,width=3cm}}} &
  \centerline{\raisebox{-0.48cm}{\psfig{bbllx=131pt,bblly=337pt,bburx=481pt,bbury=455pt,figure=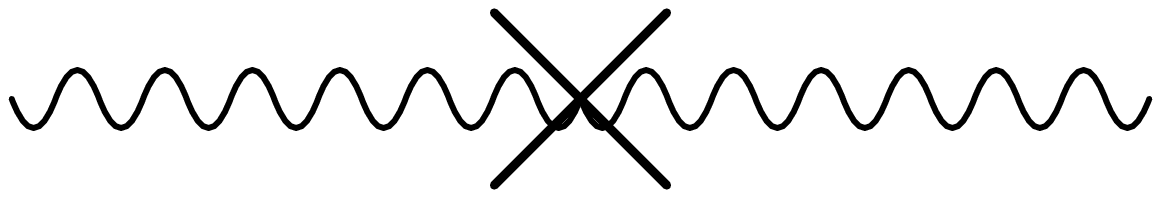,width=3cm}}} \\
  \centerline{$\hat\Sigma^{\gamma\gamma}(k^2)$} &
  \centerline{$\Sigma^{\gamma\gamma}(k^2)$} &
  \centerline{$\delta Z_2^\gamma\,k^2$}\\
  \centerline{$\hat\Sigma^{ZZ}(k^2)$} &
  \centerline{$\Sigma^{ZZ}(k^2)$} &
  \centerline{$\delta Z_2^Z (k^2-M_Z^2) -\delta M_Z^2 $} \\
  \centerline{$\hat\Sigma^{WW}(k^2)$} &
  \centerline{$\Sigma^{WW}(k^2)$} &
  \centerline{$ \delta Z_2^W (k^2-M_W^2)-\delta M_W^2$}\\
  \centerline{$\hat\Sigma^{\gamma Z}(k^2)$} &
  \centerline{$\Sigma^{\gamma Z}(k^2)$} &
  \centerline{$(\delta Z_1^{\gamma Z}
  - \delta Z_2^{\gamma Z}) M_Z^2 -\delta Z_2^{\gamma Z}\,k^2$}
\end{tabular}

\noindent
with
\begin{equation}
  \delta Z_i^{\gamma Z} = \frac{c_\theta
      s_\theta}{c_\theta^2 - s_\theta^2} 
  (\delta Z_i^Z - \delta Z_i^\gamma) \quad.
  \label{VB_Counterterme}
\end{equation}
The $\Sigma^{\alpha\beta}$ denote the unrenormalized
one-loop vector boson self energies
of the TM (see appendix). The on-shell conditions 
determine the counter terms as follows:
\begin{eqnarray}
  \delta M_W^2 &=& \Re \Sigma^{WW}(M_W^2)\nonumber\\
  \delta M_Z^2 &=&
  \Re \left[ \Sigma^{ZZ}(M_Z^2) - \frac{ (\hat\Sigma ^{\gamma Z}
    (M_Z^2))^2}{M_Z^2 + \hat \Sigma^{\gamma\gamma}(M_Z^2)} \right]
  \nonumber\\
  \delta Z_1^\gamma &=& -\Pi^\gamma(0) -
  \frac{s_\theta}{c_\theta}\, \frac{\Sigma^{\gamma Z}(0)}{M_Z^2}
  \quad \mbox{with}\quad\Pi^\gamma (0)
   = \frac{\partial \Sigma^{\gamma\gamma}}{\partial k^2} (0)
  \nonumber\\
  \delta Z_2^\gamma &=& -\Pi^\gamma(0)\nonumber\\
  \delta
  Z_1^Z &=& -\Pi^\gamma(0) -\frac{3c_\theta^2 -2s_\theta^2} {s_\theta
    c_\theta} \,\frac{\Sigma^{\gamma Z} (0)}{M_Z^2} +
  \frac{c_\theta^2-s_\theta^2}{c_\theta^2} \,\frac{\delta
    s_\theta^2}{s_\theta^2} \nonumber\\
  \delta Z_2^Z &=&
  -\Pi^\gamma(0) -2 \,\frac{c_\theta^2 -s_\theta^2} {s_\theta
    c_\theta}\, \frac{\Sigma^{\gamma Z} (0)}{M_Z^2} +
  \frac{c_\theta^2-s_\theta^2}{c_\theta^2} \,\frac{\delta
    s_\theta^2}{s_\theta^2} \, \nonumber\\ 
  \delta Z_1^W &=& -\Pi^\gamma(0) -\frac{3 - 2 s_\theta^2}{s_\theta
    c_\theta} \,\frac{\Sigma^{\gamma Z}(0)}{M_Z^2} + \frac{\delta
    s_\theta^2}{s_\theta^2}\nonumber\\
  \delta Z_2^W &=& -\Pi^\gamma(0) -2 \,\frac{c_\theta}{s_\theta}\,
  \frac{\Sigma^{\gamma Z}(0)}{M_Z^2} + \frac{\delta
    s_\theta^2}{s_\theta^2}\quad.
  \label{bos_ren_konst}
\end{eqnarray}
Herein the additional constant $\delta s_\theta^2$ appears, which is 
formally related to the $Z$-factors by
\begin{equation}
  \frac{\delta s_\theta^2}{s_\theta^2} = \frac{c_\theta}{s_\theta}\,
  (3\, \delta Z_2^{\gamma Z} - 2\, \delta Z_1^{\gamma Z}) \quad.
\end{equation}
In the SM, the renormalization of the mixing angle in the on-shell scheme
is not independent but related to the $W,Z$ mass renormalization
according to
\begin{equation}
  \frac{\delta s_W^2}{c_W^2} = \frac{\delta M_Z^2}{M_Z^2} -
  \frac{\delta M_W^2}{M_W^2} \quad.
  \label{deltas_W^2}
\end{equation}
Here, in the TM, $\delta\st^2$
has to be fixed by an extra renormalization condition. 
We do this by the identification of $\st$ with
the effective leptonic mixing angle 
$\sin\!\theta^{\rm lep}_{\rm eff}$ at the $Z$ resonance
\begin{equation}
 \st^2 \, =  \, \sin^2 \! \theta_{\rm eff}^{\rm lep} 
\end{equation}
 which determines the ratio of the leptonic
effective vector and axial vector coupling constants 
of the $Z^0$ in the following way:
\begin{equation}
  \label{zus_Ren_Bed}
  \frac{\Re(g_V^e)}{\Re(g_A^e)} = 1 - 4 s_\theta^2 \, .
\end{equation}
This is an implicit equation for  $\delta \st^2$ which enters 
ratio of the coupling constants (see equations (\ref{eff_kopp})) at 1-loop
through  the counter term to the vector form factor of the $Zee$ 
weak vertex correction. Its explicit form is given below 
in eq.~(\ref{deltastheta2}).

\medskip
The renormalized vector boson fermion vertices $\hat{\Gamma}$ are expressed
in terms of the unrenormalized vertices $\Gamma$ and the
corresponding counter terms as follows:
\hfill\\[1cm] 
 \begin{tabular}{p{3.3cm}@{\hspace{0.5cm}=\hspace{0.5cm}}p{3.3cm}
    @{\hspace{0.5cm}+\hspace{0.5cm}}p{6cm}}
  \centerline{\raisebox{-0.89cm}{\psfig{bbllx=131pt,bblly=279pt,bburx=470pt,bbury=513pt,figure=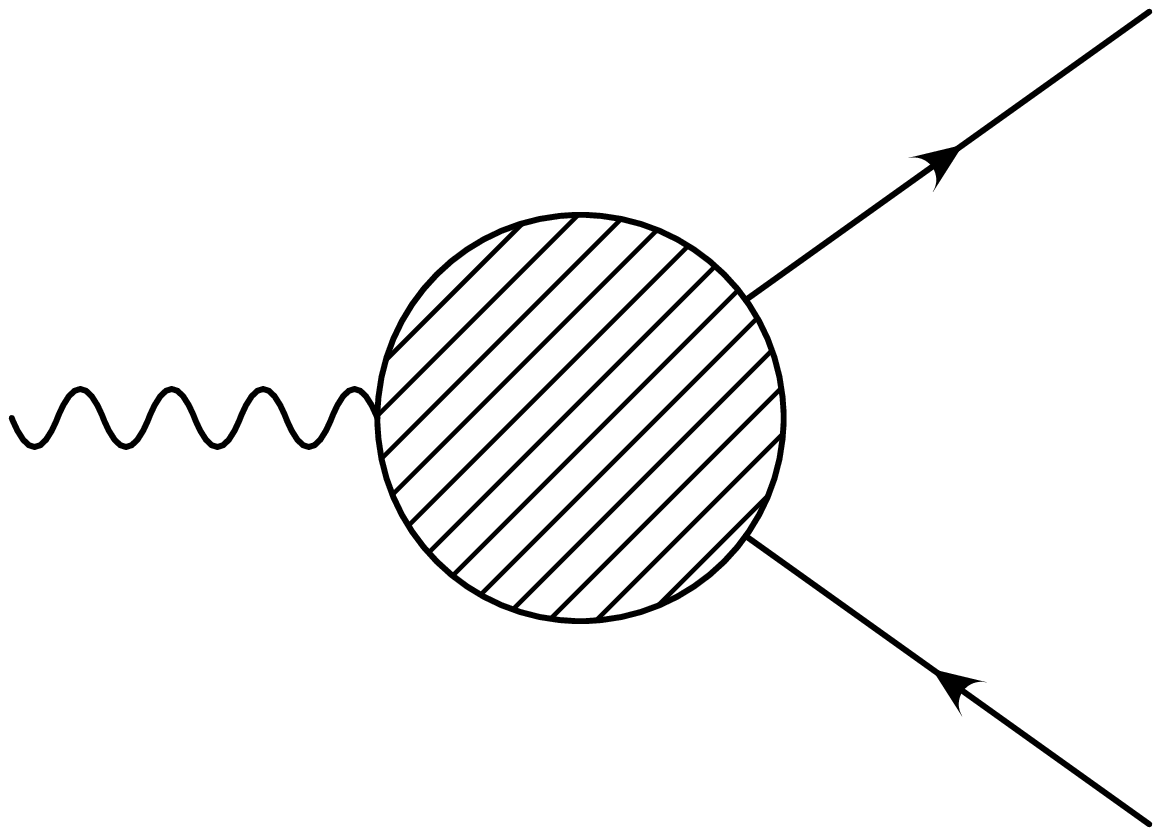,width=3cm}}} &
  \centerline{\raisebox{-0.89cm}{\psfig{bbllx=131pt,bblly=279pt,bburx=470pt,bbury=513pt,figure=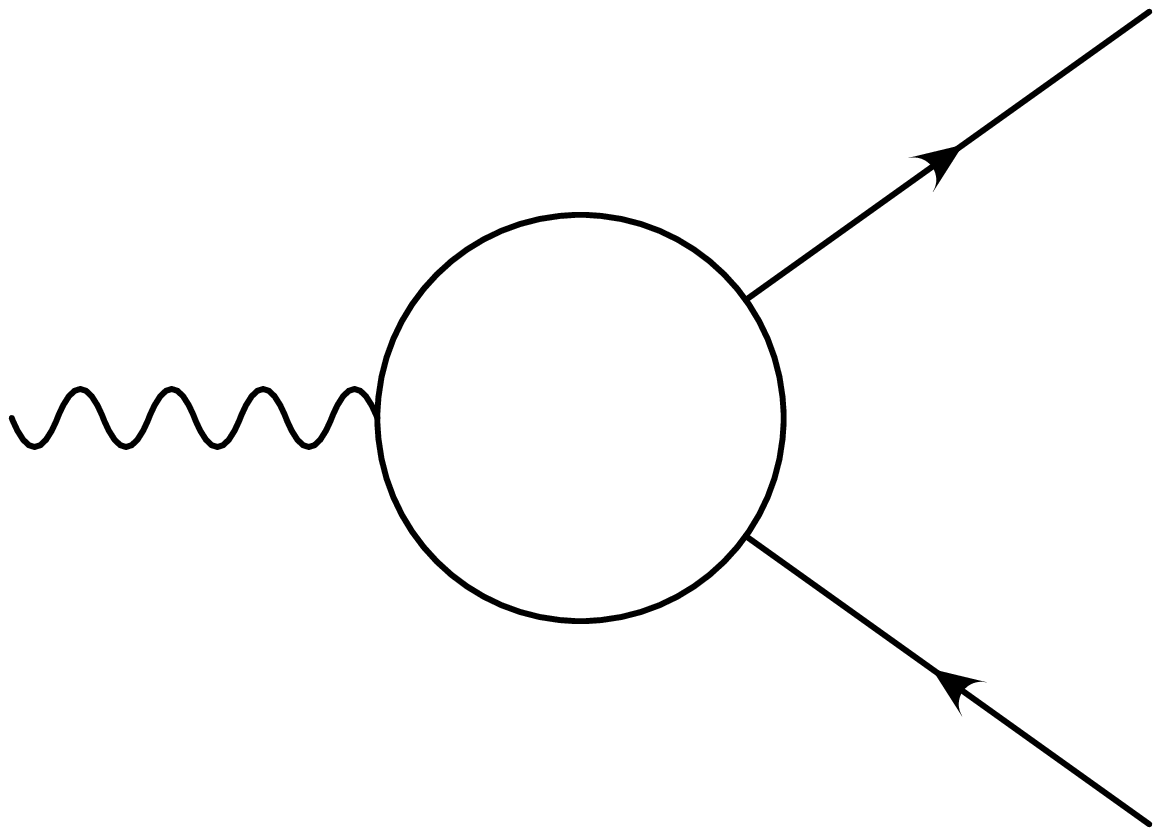,width=3cm}}} &
  \centerline{\raisebox{-0.89cm}{\psfig{bbllx=131pt,bblly=279pt,bburx=470pt,bbury=513pt,figure=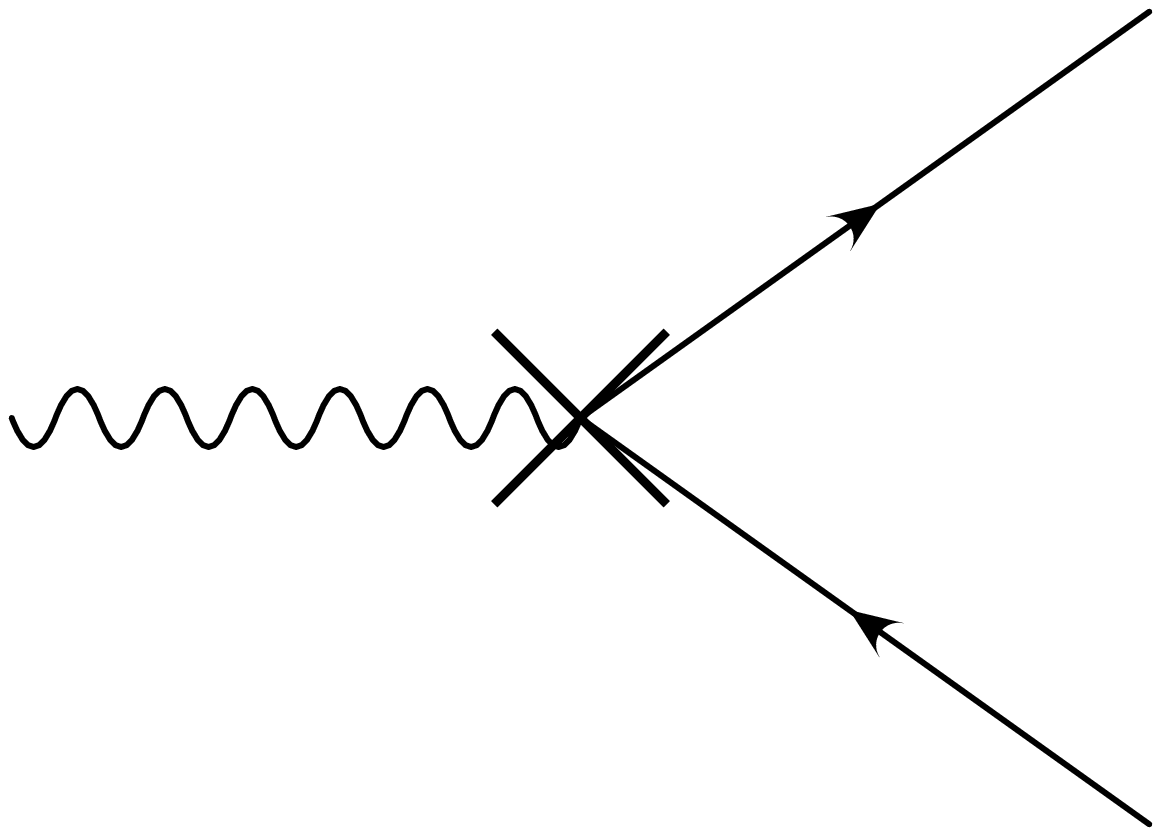,width=3cm}}}\\
  \centerline{$\hat\Gamma_{\mu}^{Zff}$} &
  \centerline{$\Gamma_{\mu}^{Zff}$} &
  \centerline{$i \frac{\T e}{\T 2 s_\theta c_\theta}
  \;\gamma_\mu \;C_V^{Zff} - i \frac{\T e}{\T 2 s_\theta c_\theta}
  \;\gamma_\mu\gamma_5 \;C_A^{Zff}$} \\
  \centerline{$\hat\Gamma_{\mu}^{W\tilde{f}f}$} &
  \centerline{$\Gamma_{\mu}^{W\tilde{f}f}$} &
  \centerline{$i \frac{\T e}{\T 2 \sqrt 2 \st} \; \gamma_\mu (1-\gamma_5)
  \;C_L^{W\tilde{f}f}$}
\end{tabular}
\begin{eqnarray}
  \mbox{with}\quad C_V^{Zff}&=&v_f(\delta Z_1^Z-\delta Z_2^Z)+ 2
  s_\theta c_\theta Q_f (\delta Z_1^{\gamma Z}-\delta Z_2^{\gamma
    Z})\nonumber\\
  &&+(v_f \delta Z_V^f + a_f \delta Z_A^f)\nonumber\\
  C_A^{Zff}&=& a_f (\delta Z_1^Z - \delta Z_2^Z)+(v_f \delta Z_A^f +
  a_f \delta Z_V^f)\nonumber\\
  C_L^{W\tilde{f}f}&=& 
  \delta Z_1^W-\delta Z_2^W+\delta Z_L  \nonumber \\ \nonumber \\
  \mbox{and}\quad v_f&=&I_3^f-2s_\theta^2 Q_f \quad,\quad a_f=I_3^f \quad.
\label{vertexct}
\end{eqnarray} 

\smallskip
\noindent
The fermion wave function renormalization constants $\delta Z_{V,A}$, and
$\delta Z_L$ resp., follow in the usual way from the ``residue = 1''
condition for the fermions attached to the vertex
(see appendix, eq.~(\ref{zfermion}) ).

\medskip
Neglecting the small terms proportional to the fermion masses $m_f$, the
$Z$ vertices have only vector and axial vector contributions:
$\hat\Lambda_{A,V}^{Zff}$ is the renormalized vector or axial-vector
correction, as it appears in the decomposition of the $Zff$ vertex
function 
\begin{equation}
  \hat\Gamma_\mu^{Z f f} = i\, \frac{e}{2 \st \ct}\, \left[ \gamma_\mu
  (v_f - a_f\gamma_5) +
  \gamma_{\mu} \hat\Lambda_V^{Z f f} +
  \gamma_{\mu} \gamma_5 \hat\Lambda_A^{Z f f}
  \right]\quad.
\label{renvertex}
\end{equation}

\medskip
\noindent Using the formulas for the effective $Z$ couplings
(\ref{eff_kopp}), the renormalization condition 
for $\st^2$, eq.~(\ref{zus_Ren_Bed}), can be written as follows:
\begin{equation}
  \Re\left\{-\frac{\hat\Pi^{\gamma Z}(M_Z^2)}{v_e}  +
    \frac{1}{2\st\ct}
    \left(
      \frac{\hat\Lambda_V^{Zee}(M_Z^2)}{v_e} -
      \frac{\hat\Lambda_A^{Zee}(M_Z^2)}{a_e} 
    \right)
  \right\}= 0 \quad,
\end{equation}
which can be solved for $\delta\st^2$ yielding
\begin{equation}
  \frac{\delta\st^2}{\st^2} = \Re
  \left\{\frac{\ct}{\st}
    \left[\frac{v_e^2-a_e^2}{a_e}\Sigma_A^e(m_e^2) +
      \frac{\Sigma^{\gamma Z}(M_Z^2)}{M_Z^2} - \frac{v_e}{2\st\ct}
      \left(\frac{\Lambda_V^{Zee}(M_Z^2)}{v_e} -
        \frac{\Lambda_A^{Zee}(M_Z^2)}{a_e} 
      \right)
    \right]
  \right\}\quad.
  \label{deltastheta2}
\end{equation}
Therein, $\Lambda_{V,A}^{Zee}$ are the vector and
axial vector form factors of the unrenormalized 1-loop $Zee$ vertex
correction in the normalization of eq.~(\ref{renvertex}),
and $\Sigma_A^e$ is the 
axial part of the $e$ self energy. 

\medskip 
In contrast to the mixing angle counter term in the minimal model,
there is no quadratic $m_t$-dependence in $\delta\st^2$. The top mass
enters via $\Sigma^{\gamma Z}$, where the dependence is only logarithmic.

\section{Radiative corrections for precision observables}
\label{radi_corr}
\noindent 
In order to fix the free parameters of the model we choose as
precise input quantities as usual the electromagnetic fine structure
constant $\alpha$ (together with the fermionic vacuum polarization
at the $M_Z$ scale), the Fermi constant $G_{\mu}$, and the $Z$ mass
$M_Z$, together with the experimental value of $\st^2$ as the fourth
input parameter for the TM. The parameters appearing in 1-loop order
are the top mass and the masses
$M_{H^0}, M_{K^0}, M_{H^\pm}$ of the standard and non-standard Higgs 
bosons. The $W$ mass $M_W$ and the $Z$ resonance parameters then follow
as predictions and can be compared with the experimental results.
\subsection{Muon decay width and $M_W$}
\noindent
The muon decay width reads in the Fermi model
\begin{equation}
  \Gamma_\mu^F=\frac{G_\mu^2 m_\mu^5}{192\, \pi^3} 
  \left(1 - \frac{8 m_e^2}{m_\mu^2}
  \right) \cdot C^{Fermi}_{QED}\quad .
  \label{fermi-Breite}
\end{equation}
In the TM it is given by the expression (see also \cite{Sirlin:78})
\begin{equation}
  \Gamma_\mu=\;\frac{\alpha^2}{384\, \pi}\; \frac{m_\mu^5}{M_W^4
    s_\theta^4} \;
  \left( 1 - \frac{8 m_e^2}{m_\mu^2} 
  \right) \cdot \left( \frac{1}{1-\Delta \tilde{r}} \right) ^2 \cdot
  C^{Fermi}_{QED}
  \label{TM-Breite}
\end{equation}
\begin{eqnarray}
  \mbox{with}&&  
  \Delta \tilde{r} = \frac{\hat\Sigma^{WW}(0)}{M_W^2} + \frac{\alpha}
  {4 \pi s_\theta^2} 
  \left(6 + \frac{10-10s_\theta^2-3\,
      (\frac{R}{c_\theta^2})\,(1-2 s_\theta^2)}{2 (1-R)} \,\ln R
  \right)\label{delta_r}\\
  \mbox{and}&&R=\frac{M_W^2}{M_Z^2}\quad.
\end{eqnarray}
The QED correction factor $C^{Fermi}_{QED}$ \cite{sirlin56}
is the same in both models.
The relation between the $W$
mass and the basic input quantities is thus given by
\begin{eqnarray}
  M_W^2&=& \rho \,M_Z^2 (1-s_\theta^2)
  \label{M_W-Best.gl.}\\
   \rho &=&\frac{A}{M_Z^2(1-\Delta \tilde{r})\,  s_\theta^2
  (1-s_\theta^2) }
  \label{c_delta-Best.gl.}\\
  \mbox{with}&&A=\frac{\alpha \pi}{\sqrt{2} G_\mu}\nonumber\quad.
\end{eqnarray}
Through $\Delta\tilde r (M_Z,M_W,\st ; m_t,M_{H^0},M_{K^0},M_{H^\pm})$
 the relations (\ref{M_W-Best.gl.}) and
(\ref{c_delta-Best.gl.}) are implicit 
equations, which can be solved iteratively
for $M_W$ and $\rho$.

\subsection{Effective $Zff$ couplings and $Z$ resonance observables}
\noindent
Having determined $\rho$ and $\Delta \tilde r$
with the help of $G_{\mu}$
 in the way described
above,
the effective couplings of the $Z$-boson to
fermions $f\neq t$
can be written in the following way:
\begin{eqnarray}
  g_V^f &=&
  \left(\rho\, \frac{1-\Delta\tilde{r}}{1+\hat\Pi^Z(M_Z^2)} 
  \right)^{\frac{1}{2}}\cdot
  \left[v_f+2\st\ct Q_f \hat\Pi^{\gamma Z} (M_Z^2) + F_V^{Zf}(M_Z^2)
  \right]\nonumber\\
  g_A^f &=&
  \left(\rho\, \frac{1-\Delta\tilde{r}}{1+\hat\Pi^Z(M_Z^2)}
  \right)^{\frac{1}{2}}\cdot
  \left[a_f + F_A^{Zf}(M_Z^2)
  \right]\quad.
  \label{eff_kopp}
\end{eqnarray}
The equations (\ref{eff_kopp}) include
besides the renormalized vertex form factors 
$ F_{V,A}^{Zf} = \hat{\Lambda}_{V,A}^{Zff} $
 the correction to the $Z$ 
propagator
\begin{eqnarray}
    \hat\Pi^Z(M_Z^2) &=& 
  \left. \Re \frac{d\hat\Sigma^Z(s)}{ds}
  \right| _{s=M_Z^2} 
\end{eqnarray}
with 
\begin{equation}
   \hat{\Sigma}^Z(s)\, = \, 
  \hat{\Sigma}^{ZZ}(s) - \frac{ (\hat\Sigma ^{\gamma Z}
    (s))^2}{s + \hat \Sigma^{\gamma\gamma}(s)} \, ,  \nonumber
\end{equation}
and the photon-$Z$ mixing
\begin{equation}  
  \hat\Pi^{\gamma Z} (M_Z^2) = \frac{\hat\Sigma^{\gamma Z}  
    (M_Z^2)}{M_Z^2 + \hat\Sigma^{\gamma\gamma}(M_Z^2)}\quad. 
\end{equation}
The self energies are from section 2.

\bigskip \noindent 
The effective coupling constants (real parts only) determine the 
on-resonance asymmetries via the combinations
   \beq
    A_f = \frac{2g_V^f g_A^f}{(g_V^f)^2+(g_A^f)^2}  \, .
\eeq
In particular: \\ 
$-$ the forward backward asymmetries
\beq A_{FB}^f =\frac{3}{4} A_e \cdot A_f \label{A_FB_had}\eeq
$-$ the left-right asymmetry
\beq A_{LR} = A_e \eeq
$-$ the $\tau$ polarization
\beq P_{\tau} = A_{\tau} \, . \eeq

 \medskip \noindent
 The fermionic partial
widths,
expressed in terms of the effective coupling constants
read up to 2nd order in the fermion masses:
\bea
\Gamma_f
  & = & \G_0
 \, \left(
     (g_V^f)^2  +
     (g_A^f)^2 (1-\frac{6m_f^2}{\mz} )
                           \right)
 \cdot   (1+ Q_f^2\, \frac{3\al}{4\pi} ) 
          + \Delta\G^f_{QCD}
\label{part_Z_Width}
\eea
with
$$
\G_0 \, =\,
  N_C^f\,\frac{\sqrt{2} G_{\mu} M_Z^3}{12\pi},
 \;\;\;\; N_C^f = 1
 \mbox{ (leptons)}, \;\; = 3 \mbox{ (quarks)}.
$$
and the QCD corrections  $ \Delta\G^f_{QCD} $
 for quark final states.
The QCD correction for the light quarks
with $m_q\simeq 0$ is given by
\beq
 \Delta\G^f_{QCD}\, =\, \G_0
  \left( (g_V^f) ^2+ (g_A^f)^2 \right)
 \cdot K_{QCD}
\eeq
with \cite{Chetyrkin:79}
$$
K_{QCD}  =   \frac{\al_s}{\pi} +1.41 \left(
  \frac{\al_s}{\pi}\right)^2 -12.8 \left(
  \frac{\al_s}{\pi}\right)^3
  -\frac{Q_f^2}{4}\frac{\al\al_s}{\pi^2} \,  .  
$$
For $b$ quarks
the QCD corrections are different due to  finite $b$ mass terms
and to top quark dependent 2-loop diagrams
 for the axial part:
\beq
 \Delta\G_{QCD}^b  =
 \Delta\G_{QCD}^d \, +\, \G_0 \left[
           (g_V^b)^2  \, R_V \,+\,
           (g_A^b)^2 \, R_A  \right]  \, . \nn
\eeq
For the coefficients $R_{V,A}$ see e.g.\ \cite{qcdb}.

\section{Results and discussion}
\label{results}
\noindent
Besides the standard input data points 
 $G_\mu = 1.16639 \cdot 10^{-5}\;{\rm GeV}^{-2}$
\cite{PDG:94}, 
$\alpha(M_Z) = 1/128.89\pm 0.09$ \cite{jegerlehner}
and $M_Z =
91.1863 \pm 0.0020\;{\rm GeV}$ \cite{Blondel:96}, we use the
effective mixing angle
$\st^2 = 0.23165\pm 0.00024$  at $M_Z$ 
as given in \cite{Blondel:96}.
Besides $m_t$, the predictions in the TM depend on the masses of the
various Higgs bosons. In general, the dependence on the Higgs masses is 
very smooth.
 In order to visualize the different dependence of
the predictions on the top mass in the various models, we display the 
results over a large top mass range and indicate the experimental data.

\subsection{The $W$ mass and the $\rho$ parameter}
\begin{figure}[htbp]
    \centerline{\psfig{figure=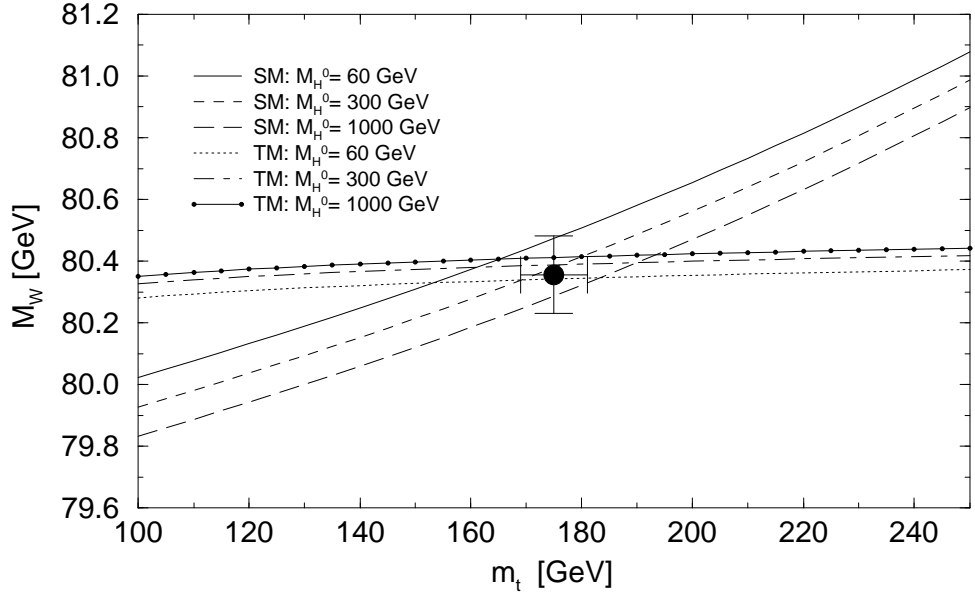,width=14cm,bbllx=18pt,bblly=55pt,bburx=552pt,bbury=360pt}}
  \caption{Top mass dependence of $M_W$ in the
    SM and the TM for various doublet Higgs masses $M_{H^0}$. The input
    values for the TM Higgs masses $M_{K^0}$ and $M_{H^\pm}$ are 300 GeV.}
  \label{fig:tr_m}
\end{figure}

\noindent In Figure \ref{fig:tr_m} the top mass dependence of
 $M_W$ is displayed 
for a set of masses for the doublet Higgs boson $H^0$, both
in the minimal model (SM) and the standard model with the extra
triplet (TM). The other Higgs masses have been fixed at 300 GeV.
The dependence on both $m_t$ and $M_{H^0}$ is weaker in the TM compared to
the SM.
The experimental result 
\begin{equation}
  M_W = 80.356 \pm 0.125\;\mbox{GeV}\quad\cite{Blondel:96}\quad,\quad
  m_t = 175 \pm 6\;\mbox{GeV}\quad\cite{Tipton:96}\nonumber
\end{equation}
is shown as the data 
point with error bars. 
It is placed right in the cross-over region of the two models.

\begin{figure}[htbp]
    \centerline{\psfig{figure=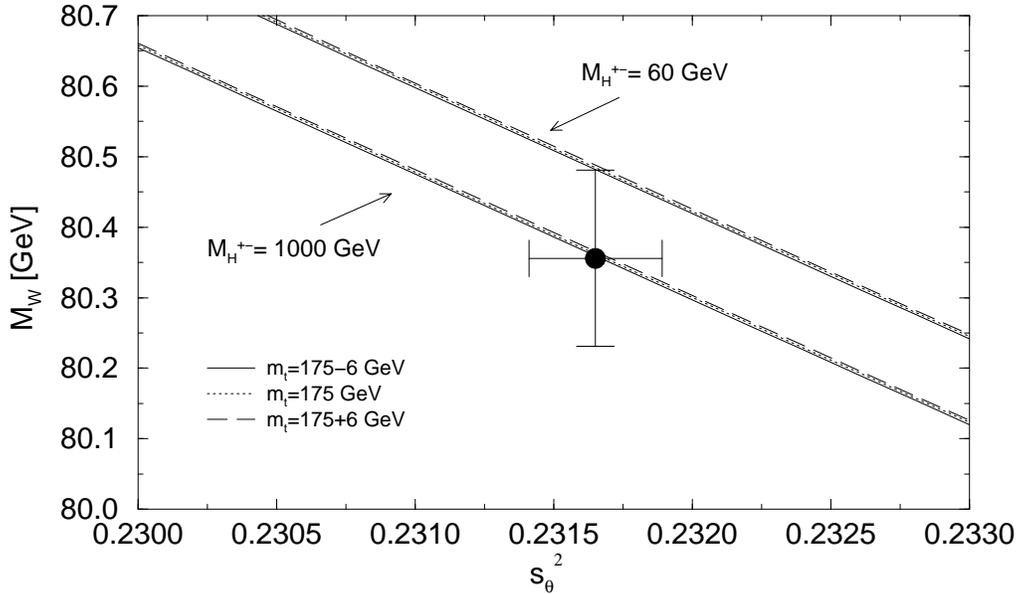,width=14cm,bbllx=18pt,bblly=55pt,bburx=552pt,bbury=360pt}}
  \caption{Dependence of $M_W$ on the input parameter
    $\st^2$ for various values of $m_t$ and $M_{H^\pm}$
    in the TM. The masses
    for the neutral Higgs bosons are fixed at 300 GeV.}
  \label{fig:tr_o}
\end{figure}

In the TM, $M_W$ has a strong dependence on the value of the input
parameter $\st^2$. This is illustrated in Fig.~\ref{fig:tr_o} 
for different
values of the charged Higgs mass, with both the neutral Higgs masses
at 300 GeV. Compared with the experimental data, the sensitivity 
to $M_{H^\pm}$ is not very striking. Higher masses are slightly prefered,
in particular for a low value of $\st^2$. The variation with $m_t$
in its experimental $1\sigma$ range is hardly visible.

\begin{figure}[htbp]
    \centerline{\psfig{figure=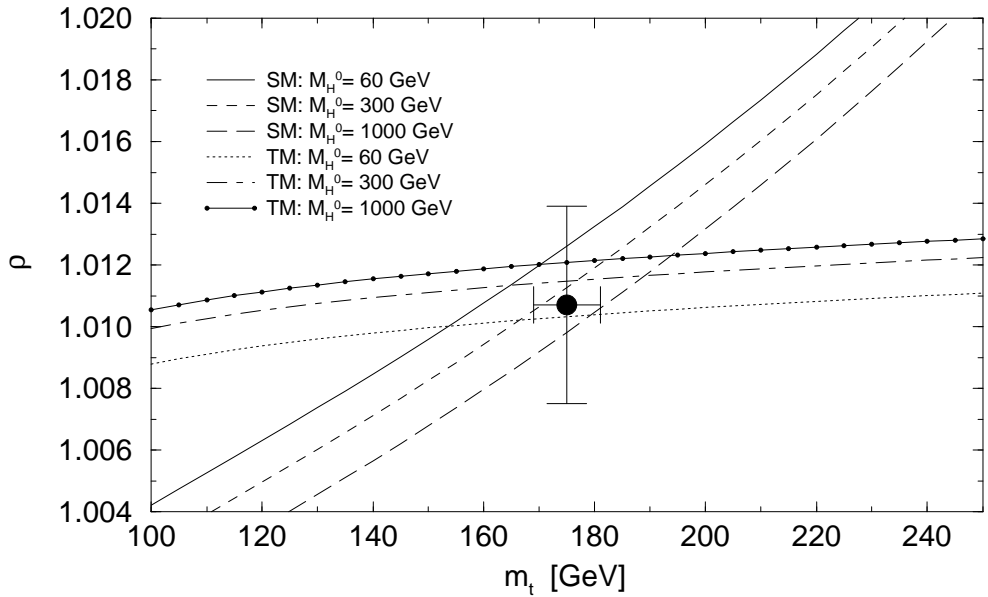,width=14cm,bbllx=18pt,bblly=55pt,bburx=552pt,bbury=360pt}}
  \caption{Top mass dependence of the $\rho$
    parameter in the SM  and the TM for various doublet Higgs masses
    $M_{H^0}$. The input values for the TM Higgs masses $M_{K^0}$ and
    $M_{H^\pm}$ are 300 GeV.}
  \label{fig:tr_xa}
\end{figure}

\begin{figure}[htbp]
    \centerline{\psfig{figure=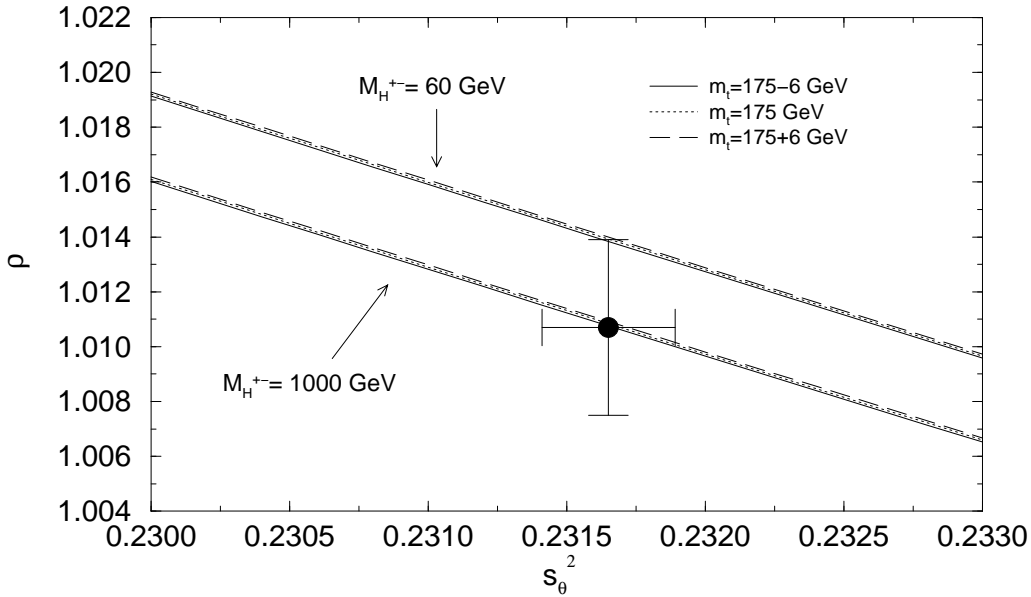,width=14cm,bbllx=18pt,bblly=55pt,bburx=552pt,bbury=360pt}}
  \caption{Dependence of the $\rho$ parameter on the 
    input parameter $\st^2$ in the TM for various values of $m_t$ and
    $M_{H^\pm}$. The masses of the neutral Higgs bosons are fixed
    at 300 GeV.}
  \label{fig:tr_xc}
\end{figure}

An interesting quantity is the $\rho$-parameter, eq.~(\ref{rho-cdelta})
which can act as an indicator for a deviating Higgs structure.
Since also in the SM $\rho$ is different from unity by radiative corrections,
a sensible comparison of different models is only possible at the 
1-loop order.  The experimental value
derived from $M_W$, $M_Z$ and
$\sin^2\!\theta_{\rm eff}^{\rm lep}$ \cite{Blondel:96} is given by
\begin{equation}
  \rho = 1.0107 \pm 0.0032.
\end{equation}
The dependence of $\rho$ on the model parameters is shown in
Fig.~\ref{fig:tr_xa} and Fig.~\ref{fig:tr_xc} for SM and TM, together
with the experimental data.  The models overlap in the region of the
data, which is equivalent to the situation in the corresponding figure
with $M_W$.

\subsection{$Z$ boson observables}
\noindent Precision observables at the $Z$ resonance are the total
and partial $Z$ decay widths and the peak asymmetries.
The total $Z$ width can be
expressed as the sum of the fermionic partial widths
\begin{equation}
  \Gamma_Z = \sum_f \Gamma_f\quad,
\end{equation}
which are defined in equation (\ref{part_Z_Width}).

\begin{figure}[htbp]
    \centerline{\psfig{figure=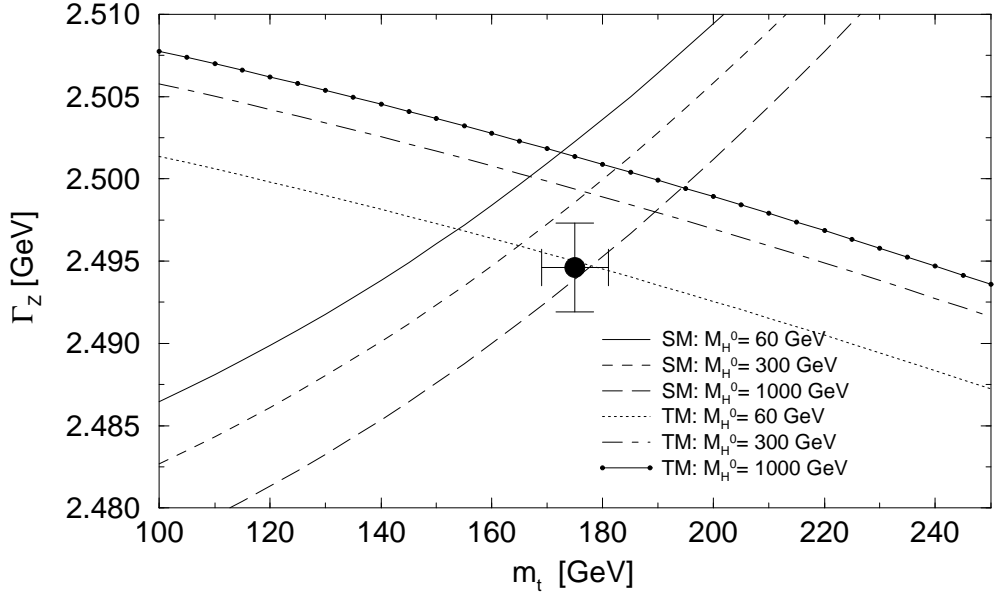,width=14cm,bbllx=18pt,bblly=55pt,bburx=552pt,bbury=360pt}}
  \caption{Top mass dependence of the total $Z$
    width in the SM and the TM for various doublet Higgs masses
    $M_{H^0}$. The input values for the TM Higgs masses $M_{K^0}$ and
    $M_{H^\pm}$ are 300 GeV.}
  \label{fig:tr_g}
\end{figure}

\begin{figure}[htbp]
    \centerline{\psfig{figure=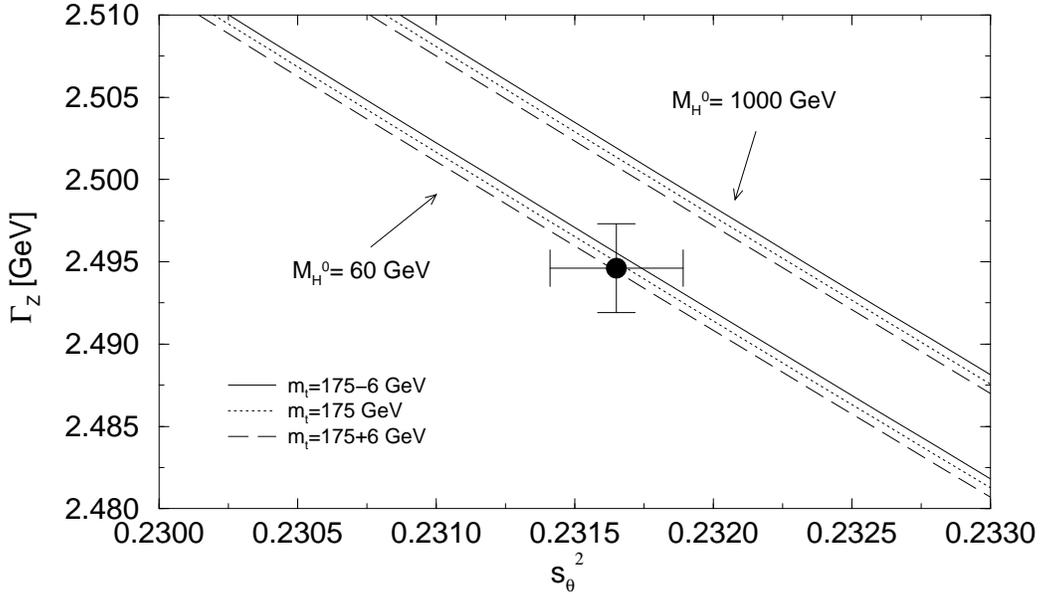,width=14cm,bbllx=18pt,bblly=55pt,bburx=552pt,bbury=360pt}}
  \caption{Dependence of the total $Z$ width on the 
    input parameter $\st^2$ for various values of $m_t$ and
    $M_{H^0}$. The masses of the triplet Higgs bosons are fixed at
    300 GeV.}
  \label{fig:tr_i}
\end{figure}

Similar to $M_W$, we display the total width $\G_Z$ in 
Fig.~\ref{fig:tr_g}
versus $m_t$  for the SM and the TM, 
together with the experimental data point
$\Gamma_Z = 2.4946 \pm 0.0027\;{\rm GeV}$ \cite{Blondel:96}.
Although the models show a different behaviour with $m_t$ and $M_H$,
they coincide in the region where both models agree with the data.
It is interesting to note that the SM has a preference for a heavy Higgs
from the observable $\G_Z$, whereas the mixing angle measurement 
requires a light Higgs boson. In the TM, a light $H^0$ is compatible 
with all precision observables.
Fig.~\ref{fig:tr_i} makes the TM correlation between $\st^2$ and $M_{H^0}$
in the $Z$ width more explicit for the measured value of the top mass.

\begin{figure}[htbp]
  \centerline{\psfig{figure=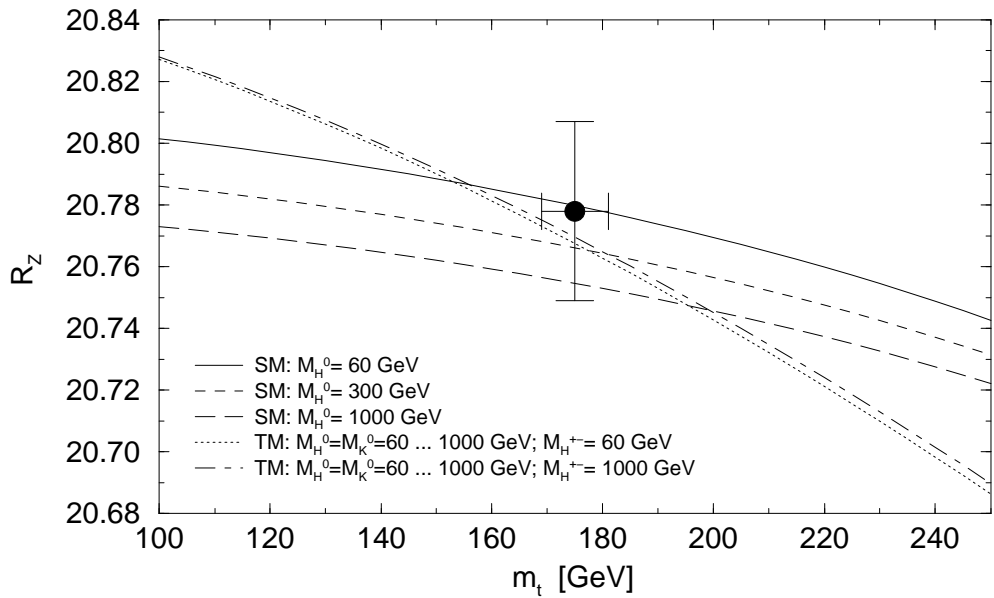,width=14cm,bbllx=18pt,bblly=55pt,bburx=552pt,bbury=360pt}}
  \caption{Top mass dependence of $R_Z$ in the
    SM and the TM for various Higgs masses.}
  \label{fig:tr_a}
\end{figure}

Partial widths are conveniently discussed in terms of the ratios
\begin{equation}
  R_Z = \frac{\Gamma_{\rm had}}{\Gamma_e}\quad,\quad
  R_c = \frac{\Gamma_c}{\Gamma_{\rm had}}\quad\mbox{and}\quad
  R_b = \frac{\Gamma_b}{\Gamma_{\rm had}}\quad,
  \label{R_ZR_cR_b}
\end{equation}
which are experimentally determined to \cite{Blondel:96}
\begin{eqnarray}
  R_Z &=& 20.778 \pm 0.029\nonumber\\
  R_c &=& 0.1715 \pm 0.0056\nonumber\\
  R_b &=& 0.2178 \pm 0.0011\nonumber\quad.
\end{eqnarray}
The predictions for $R_Z$ by the SM and the TM are illustrated in
Fig.~\ref{fig:tr_a}. In contrast to the previously discussed
observables, the $m_t$-dependence of $R_Z$ is stronger 
in the TM. $R_Z$ is, however, completely insensitive  to any Higgs
mass. Again we encounter the situation that the two models
coincide exactly in that range where the experimental data
are placed.

\begin{figure}[htbp]
    \centerline{\psfig{figure=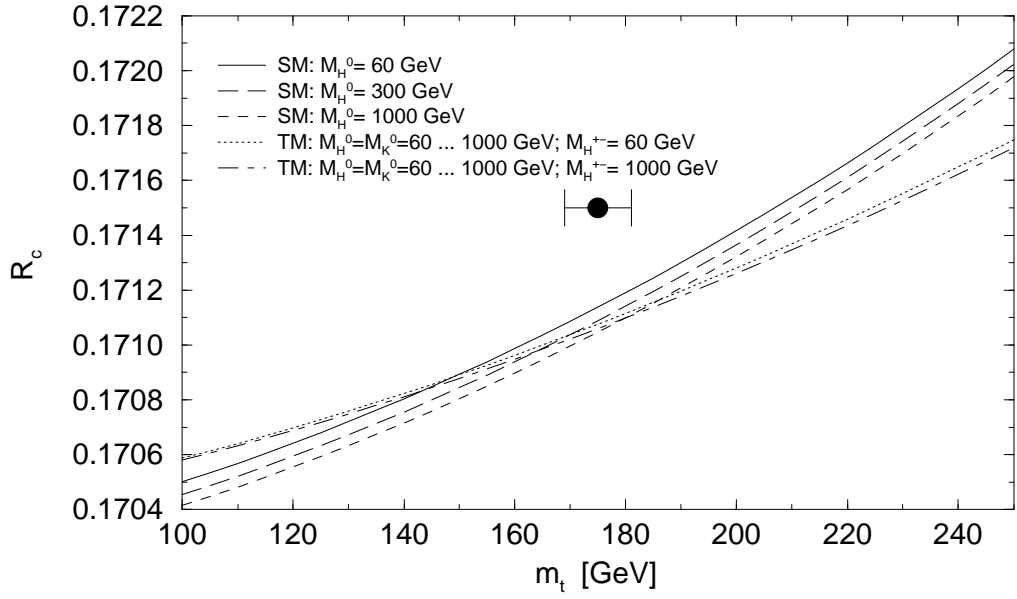,width=14cm,bbllx=18pt,bblly=55pt,bburx=552pt,bbury=360pt}}
  \caption{Top mass dependence of $R_c$ in the
    SM and the TM for various Higgs masses. The error bar of $R_c$
    covers the full vertical axis.}
  \label{fig:tr_v}
\end{figure}

The quantity $R_c$ is not very instructive with respect to the Higgs
sector. Fig.~\ref{fig:tr_v} contains the predictions for $R_c$, which in
view of the comparatively large experimental error can be considered
as identical and in best agreement with the data.

\begin{figure}[htbp]
    \centerline{\psfig{figure=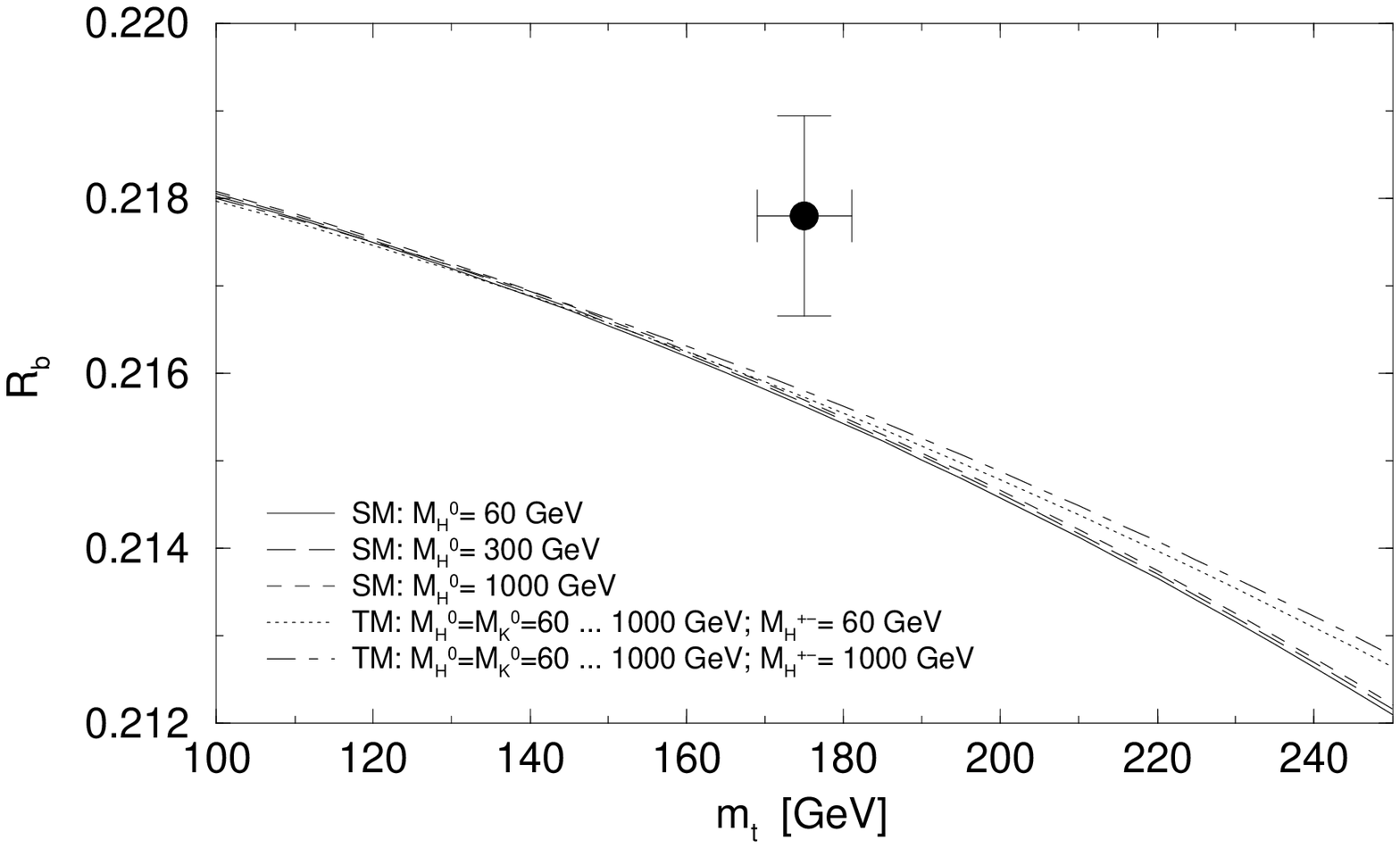,width=14cm,bbllx=18pt,bblly=55pt,bburx=552pt,bbury=360pt}}
  \caption{Top mass dependence of $R_b$ in the
    SM and the TM for various Higgs masses.}
  \label{fig:tr_c}
\end{figure}

An observable of special interest is the quantity $R_b$ with its 
experimental value about $1.8\, \sigma$ above the SM prediction.
Its special sensitivity to $m_t$ is based on the virtual presence
of the top quark in the $Z \bar b b$ vertex corrections.
Fig.~\ref{fig:tr_c} shows  the predictions of both the SM and TM,
which with exception of very high top masses are the same,
with practically no Higgs dependence. The deviation from the data
point hence is also the same in both type of models.

\begin{figure}[htbp]
  \centerline{\psfig{figure=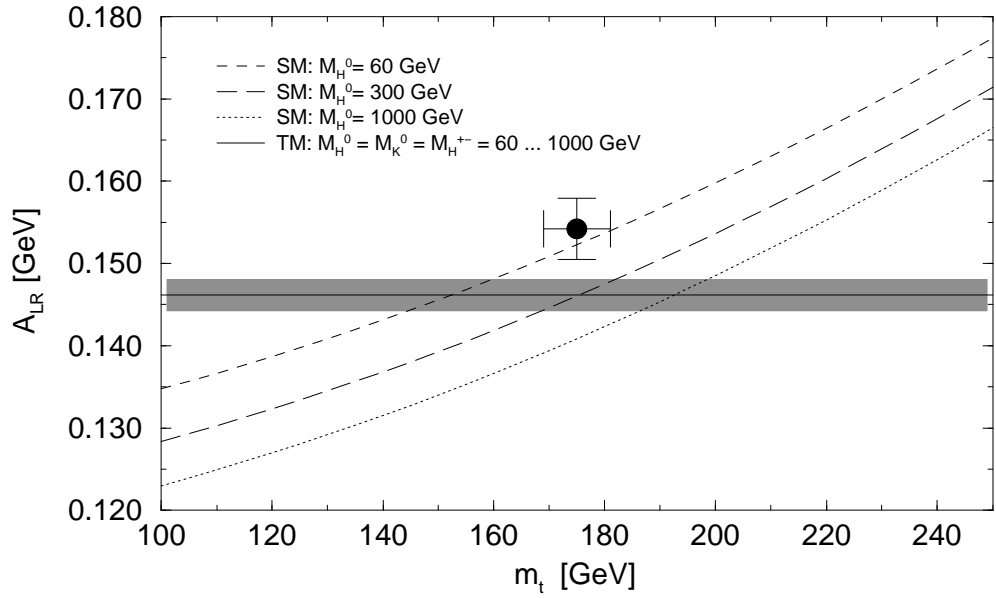,width=14cm,bbllx=18pt,bblly=55pt,bburx=552pt,bbury=360pt}}
  \caption{Left/Right asymmetry
    in the SM and the TM. The shaded area 
    corresponds to a variation of $\st^2=0.23165\pm0.00024$.}
  \label{fig:tr_xy}
\end{figure}

\medskip 
The leptonic on-resonance asymmetries are in the TM completely determined
by the value of the input parameter $\st^2$, which is the
leptonic mixing angle (and actually determined from asymmetry
measurements). For the purpose of illustration, we present in
Fig.~\ref{fig:tr_xy} the left-right asymmetry $A_{LR}$ as
predicted by the SM in terms of $m_t$ and $M_H$, and the range
corresponding to the TM input
   $\st^2 =0.023165\pm 0.00024$.
This range, indicated by the shaded area, can be identified with
the TM ``prediction''. The SM requires a light Higgs boson, which is 
disfavoured by the total width $\G_Z$ (Fig.~\ref{fig:tr_g}),
in contrast to the TM. The experimental value as measured by the SLD
collaboration is given by \cite{Blondel:96}
\begin{equation}
  A_{LR} = 0.1542 \pm 0.0037 \quad.
\end{equation}

\begin{figure}[htbp]
  \centerline{\psfig{figure=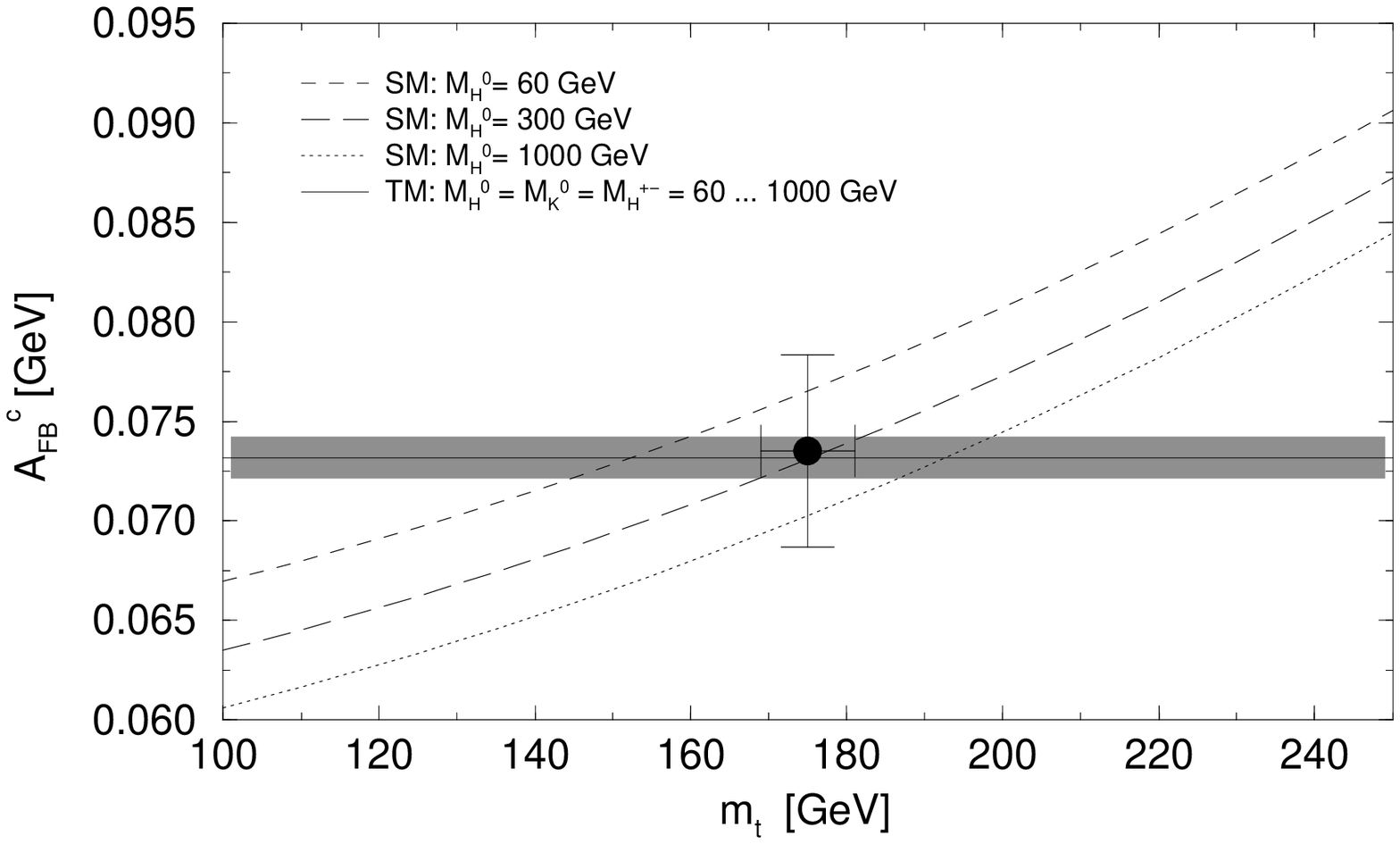,width=14cm,bbllx=18pt,bblly=55pt,bburx=552pt,bbury=360pt}}
  \caption{Forward/backward asymmetry
    for  charm quarks in the SM and the TM. The shaded area 
    corresponds to a variation of $\st^2=0.23165\pm0.00024$.}
  \label{fig:tr_xl}
\end{figure}

\begin{figure}[htbp]
  \centerline{\psfig{figure=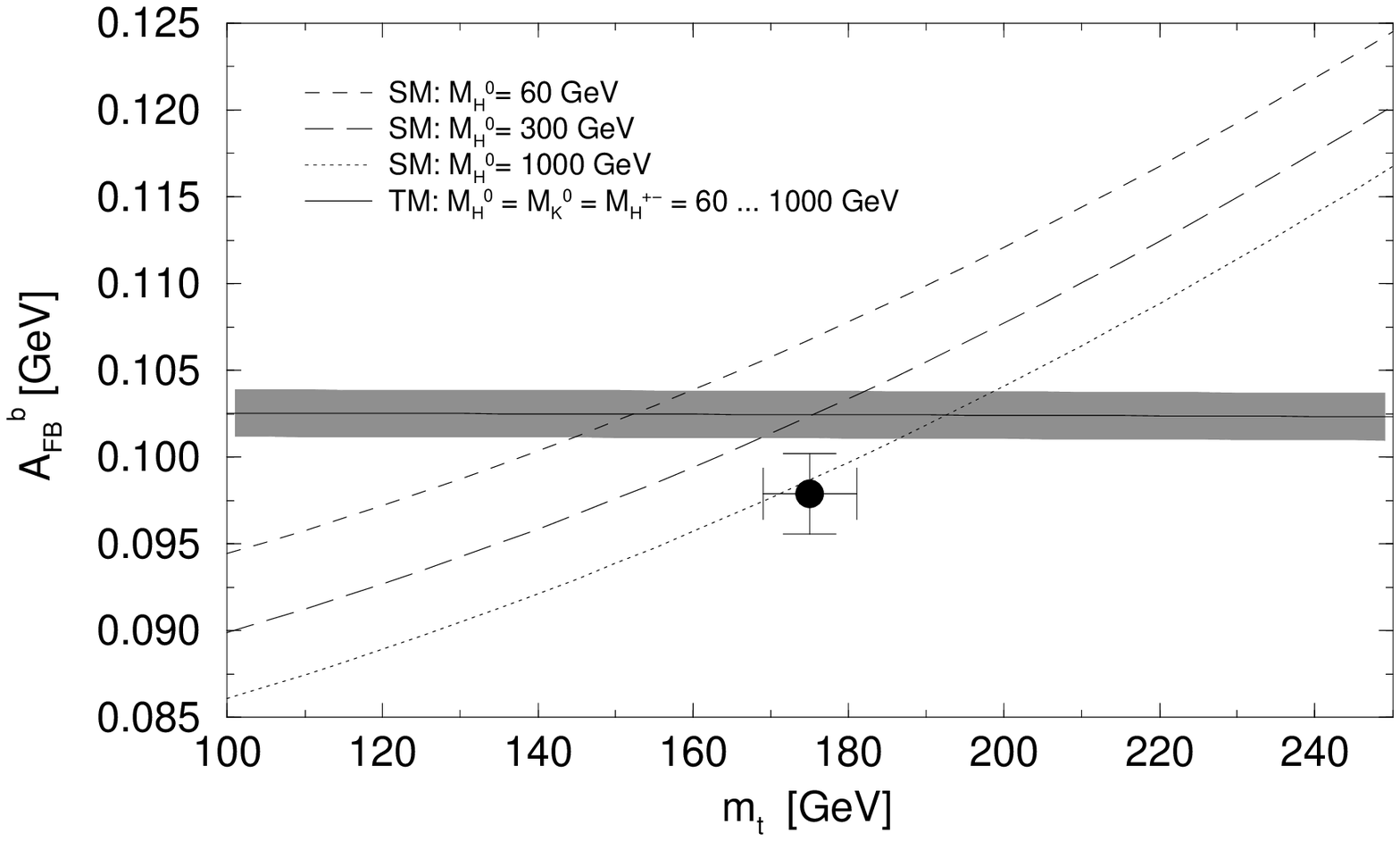,width=14cm,bbllx=18pt,bblly=55pt,bburx=552pt,bbury=360pt}}
  \caption{Forward/backward asymmetry
    for  bottom quarks in the SM and the TM. The shaded area 
   corresponds to a variation of $\st^2=0.23165\pm0.00024$.}
  \label{fig:tr_xk}
\end{figure}

The hadronic forward-backward asymmetries for $c$ and $b$ quark final 
states contain besides $A_e$ the additional factors $A_{c,b}$
in eq. (\ref{A_FB_had}). 
In practice, however, the model dependence beyond $\st^2$
cancels in the ratios. Consequently, the TM predictions in
Fig.~\ref{fig:tr_xl} and Fig.~\ref{fig:tr_xk} appear as a top and Higgs mass
independent horizontal line for each fixed value of $\st^2$.
Varying $\st^2$ in the $1\, \sigma$ range yields the shaded band.
The SM predictions, on the other hand, do depend on $m_t$ and $M_H$,
essentially through $\st^2$.
 The experimental results are given by
\cite{Blondel:96}
\begin{equation}
  A_{FB}^c = 0.07351 \pm 0.00484 \quad\mbox{and}\quad
  A_{FB}^b = 0.09790 \pm 0.00231 \quad.
\end{equation}

Whereas $A_{FB}^c$ is perfect for both models, 
$A_{FB}^b$ needs a large Higgs mass in the SM, opposite to the requirement
from $A_{LR}$. The TM coincides with the SM in the intermediate range
of $M_{H^0}$; it is also slightly higher than the experimental value.

\newpage
\section{Conclusions}
\noindent
We have presented a complete 
1-loop calculation of 
electroweak precision observables in the extension of the SM by an
extra Higgs triplet, where the $\rho$-parameter can be different
from unity already at the tree level. 
Since the gauge - fermion sector has one free parameter more compared
to the SM,  one additional data point is required for fixing the input
parameters. Choosing the effective leptonic mixing angle, 
the observables depend, besides on $\st^2$ and 
the conventional input $\alpha, M_Z, G_{\mu}, m_t$, on the 
mass of the doublet Higgs boson $H^0$ and on the masses of the extra 
non-standard Higgs bosons as free parameters.
The predictions of the SM and the TM coincide for all observables in the
experimental range of the top mass $m_t=175\pm 6$ GeV.
In this range, both models fully agree with the experimental
precision data, with two exceptions: $R_b$, $A^b_{FB}$, where
both models show similar deviations from the data.
The two types of models are thus indistinguishable, and no signal for
a non-standard Higgs structure can be found in the data.
In the TM all observables which show
a dependence on the doublet Higgs mass, are consistent with a low
value of $M_{H^0}$, whereas in the SM some observables like $\Gamma_Z$
advocate a large value for $M_{H^0}$.
\newpage

\begin{appendix}
\section*{Appendix}
\setcounter{equation}{0}
\def\theequation{A.\arabic{equation}}
\noindent This section contains the analytic expressions for the
vector boson and fermion self energies and the $Z \bar f f$ vertex
corrections with internal Higgs states. 
Only those contributions are listed which are different from the minimal 
standard model.
$\tilde{f}$ always denotes 
the isospin partner of the fermion $f$.  Moreover, the following
abbreviations are used:
\begin{equation}
  c_\delta = \cos\delta\quad,\quad s_\delta = \sin\delta \quad.
\end{equation}
The scalar 1-, 2- and 3-point functions in dimensional regularization
  are given by
\begin{eqnarray}
  \frac{i}{16 \pi^2}\, A &=& \mu^{4-D} \int \frac{d^D k}{(2\pi)^D}\;
  \frac{1}{k^2-m^2} \nonumber\\
  \frac{i}{16 \pi^2}\,B_0 &=& \mu^{4-D} \int
  \frac{d^D k}{(2 \pi)^D}\;
  \frac{1}{[k^2-m_1^2][(k+p)^2-m_2^2]}\nonumber\\
  \frac{i}{16\pi^2}\, C_0 &=& \mu^{4-D} \int \frac{d^D
    k}{(2\pi)^D}\;
  \frac{1}{[k^2-m_1^2][(k+p_1)^2-m_2^2][(k+p_1+p_2)^2-m_3^2]}\quad,
\end{eqnarray}
We also need the scalar coefficients in the  
tensor integral decompositions  
\cite{passarino-veltman}
\begin{eqnarray}
  B^\mu &= & p^\mu\: B_1(p^2,m_1,m_2)\nonumber\\
  B^{\mu\nu} &=& g^{\mu\nu} B_{22}(p^2,m_1,m_2)+ p^\mu\, p^\nu\:
  B_{21}(p^2,m_1,m_2)\nonumber\\
  C^\mu &=& p_1^\mu\: C_{11} + p_2^\mu\: C_{12}\nonumber\\
  C^{\mu\nu} &=& g^{\mu\nu} C_{20} +p_1^\mu\, p_1^\nu\: C_{21} +
  p_2^\mu\, 
  p_2^\nu\: C_{22} + (p_1^\mu\, p_2^\nu + p_1^\nu\, p_2^\mu)\:
  C_{23}\quad.
\end{eqnarray}
For the 2-point functions they are given by
\begin{eqnarray}
  B_1(p^2,m_1,m_2)&=& \frac{1}{2p^2} \left[ A(m_1)-A(m_2) +
  (m_2^2-m_1^2-p^2) B_0(p^2,m_1,m_2)\right]\nonumber\\
  B_{22}(p^2,m_1,m_2)&=& \frac{1}{6} \left[ A(m_2) + 2m_1^2
  B_0(p^2,m_1,m_2)\right.\nonumber\\ 
  &&\left.+(p^2+m_1^2-m_2^2)B_1(p^2,m_1,m_2)
  +m_1^2+m_2^2-\frac{p^2}{3}\right]\nonumber\\
  B_{21}(p^2,m_1,m_2)&=&\frac{1}{3p^2}\left[A(m_2) - m_1^2
  B_0(p^2,m_1,m_2)\right. \nonumber\\
  &&\left. -2(p^2 + m_1^2-m_2^2) B_1(p^2,m_1,m_2)
  -\frac{m_1^2+m_2^2}{2}  + \frac{p^2}{6}\right] \quad.
  \label{kov-Koeff}
\end{eqnarray}
For the corresponding 
expressions in the 3-point functions see e.g.~\cite{Hollik:93a}.\\

\newpage
\noindent
{\bf Vector boson self energies:}

\smallskip \noindent
For the vector boson self energies three diagram
topologies with internal Higgs lines contribute. The analytic 
expressions given below  
correspond to the sum over all possibilities for $S$.\\
\centerline{\psfig{figure=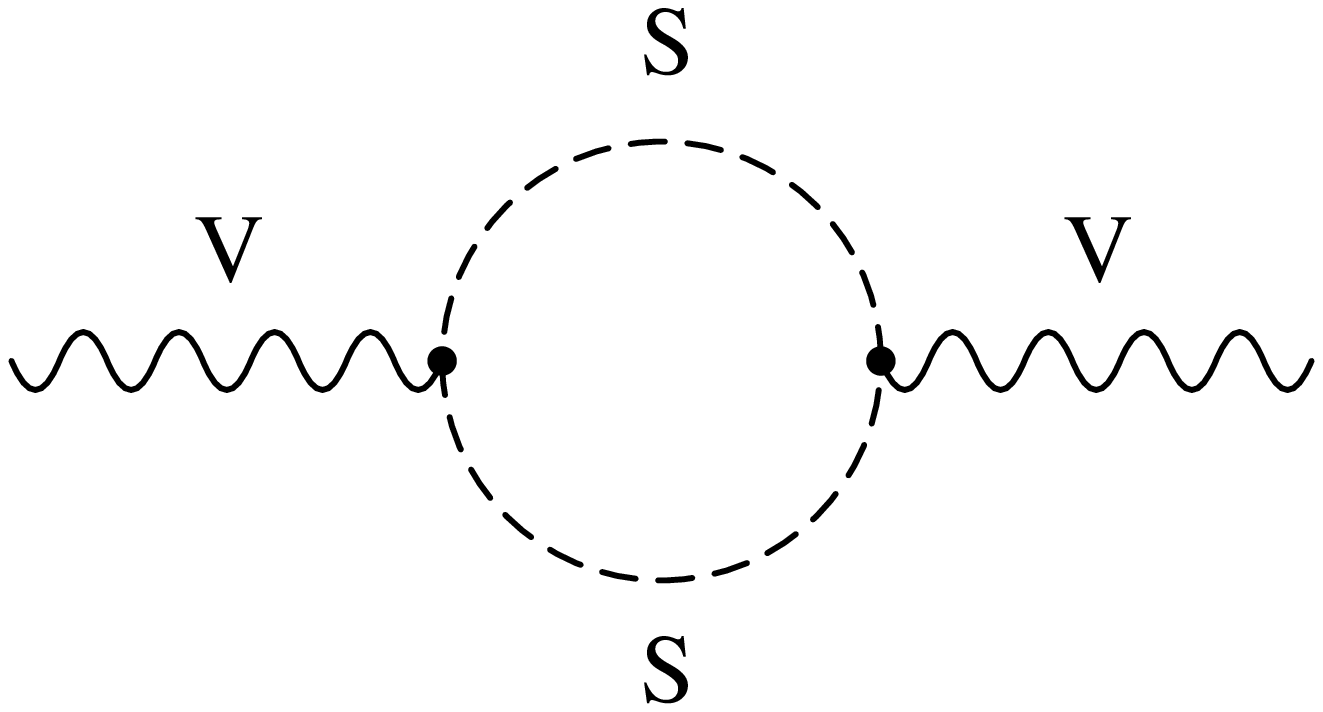,width=5cm}}
\begin{eqnarray}
  \Sigma^{\gamma\gamma}_{(SS)} &=& \frac{\alpha}{4\pi}
  \left\{ - 4 B_{22}(k^2,M_{H^\pm},M_{H^\pm}) - 4 B_{22}(k^2,M_W,M_W)
  \right\}\nonumber\\
  \Sigma^{ZZ}_{(SS)} &=& \frac{\alpha}{4\pi} 
  \left(- \frac{1}{\st^2\ct^2} 
  \right)
  \left\{ B_{22}(k^2,M_{H^0},M_Z)
    + (c_\delta^2+c_\theta^2-s_\theta^2)^2
    B_{22}(k^2,M_{H^\pm},M_{H^\pm})
  \right.\nonumber\\
  &&
  \left.
    + (s_\delta^2+c_\theta^2-s_\theta^2)^2
    B_{22}(k^2,M_W,M_W)
    + 2 s_\delta^2 c_\delta^2
    B_{22}(k^2,M_{H^\pm},M_W)
  \right\}\nonumber\\
  \Sigma^{\gamma Z}_{(SS)} &=& \frac{\alpha}{4\pi} \frac{1}{\st\ct} 
  \left\{ 2 (\cd^2-\st^2+\ct^2) B_{22}(k^2,M_\Hpm,M_\Hpm)
  \right.\nonumber\\
  &&
  \left.
    + 2 (\sd^2-\st^2+\ct^2)B_{22}(k^2,M_W,M_W)
  \right\}\nonumber\\
  \Sigma^{WW}_{(SS)} &=& \frac{\alpha}{4\pi} 
  \left( -\frac{1}{\st^2}
  \right)
  \left\{\sd^2 B_{22}(k^2,M_\Hn,M_\Hpm)
    +\cd^2 B_{22}(k^2,M_\Hn,M_W)
  \right.\nonumber\\
  &&
  \left.
    + 4 \cd^2 B_{22}(k^2,M_\Kn,M_\Hpm)
    + 4 \sd^2 B_{22}(k^2,M_\Kn,M_W)
  \right.\nonumber\\
  &&
  \left.
    + \sd^2 B_{22}(k^2,M_Z,M_\Hpm)
    + \cd^2 B_{22}(k^2,M_Z,M_W)
  \right\}\\
  \mbox{}&&\mbox{}\hspace{13cm}\mbox{}\nonumber
\end{eqnarray}
\newpage
\centerline{\psfig{figure=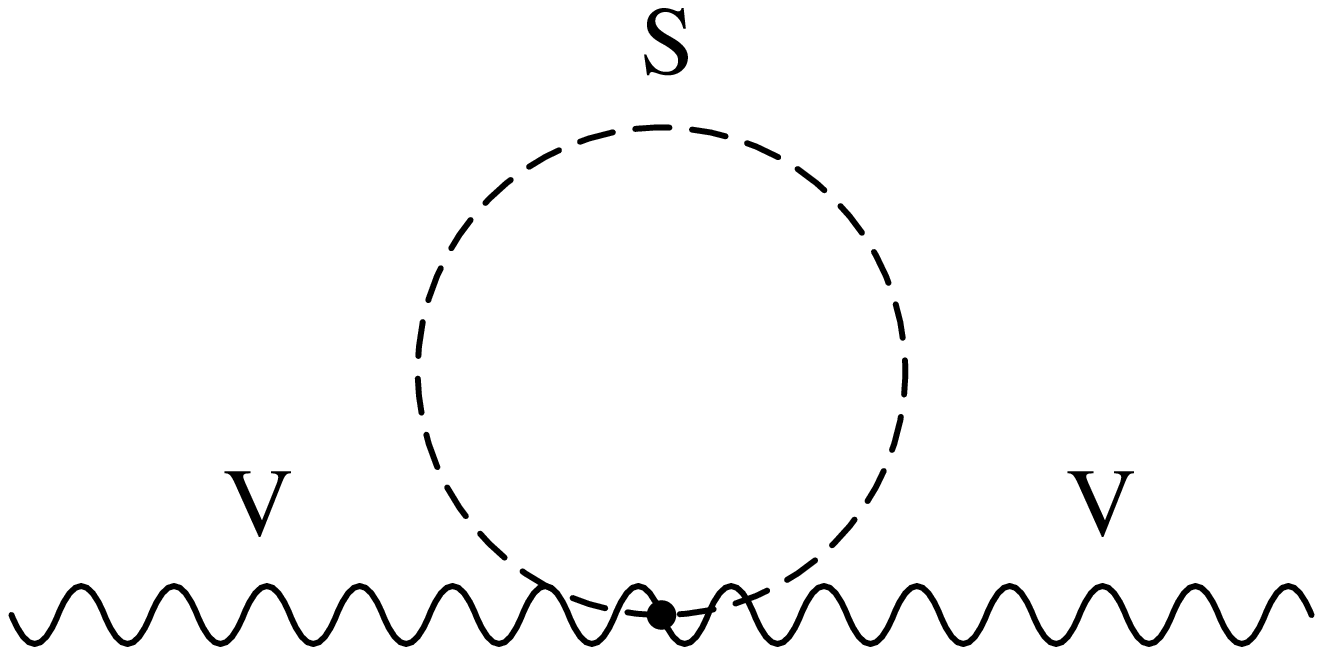,width=5cm}}
\begin{eqnarray}
  \Sigma^{\gamma\gamma}_{(S)} &=& \frac{\alpha}{4\pi}
  \left\{2 A(M_{H^\pm}) + 2 A(M_W)
  \right\}\nonumber\\
  \Sigma^{ZZ}_{(S)} &=& \frac{\alpha}{4\pi} \frac{1}{4\st^2\ct^2} 
  \left\{ A(M_\Hn) +  A(M_Z) + 
    2 ((\ct^2-\st^2+\cd^2)^2+\sd^2\cd^2) A(M_\Hpm)
  \right.\nonumber\\
  &&
  \left.
    + 2 ((\ct^2-\st^2+\sd^2)^2+\sd^2\cd^2) A(M_W)
  \right\}\nonumber\\
  \Sigma^{\gamma Z}_{(S)} &=& \frac{\alpha}{4\pi} \frac{1}{\st\ct} 
  \left\{ (\st^2-\ct^2-\cd^2)A(M_\Hpm) +
    (\st^2-\ct^2-\sd^2) A(M_W)
  \right\}\nonumber\\
  \Sigma^{WW}_{(S)} &=& \frac{\alpha}{4\pi} \frac{1}{4\st^2} 
  \left\{ A(M_\Hn)+4 A(M_\Kn) + A(M_Z) + 2 (1+\cd^2) A(M_\Hpm)
  \right.\nonumber\\
  &&
  \left.
    + 2 (1 + \sd^2) A(M_W)
  \right\}\\
  \mbox{}&&\mbox{}\hspace{13cm}\mbox{}\nonumber
\end{eqnarray}
\centerline{\psfig{figure=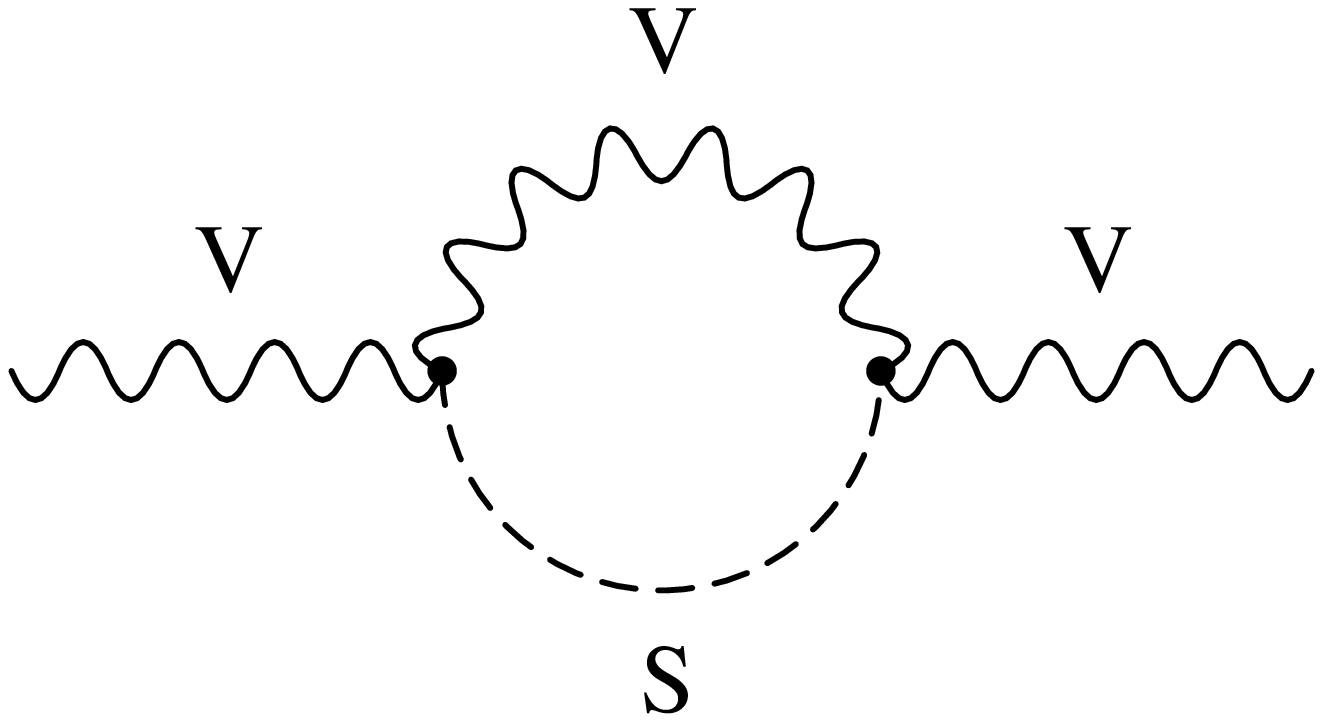,width=5cm}}
\begin{eqnarray}
  \Sigma^{\gamma\gamma}_{(VS)} &=& \frac{\alpha}{4\pi}
  2 M_W^2 B_0(k^2,M_W,M_W)\nonumber\\
  \Sigma^{ZZ}_{(VS)} &=& \frac{\alpha}{4\pi} \frac{M_Z^2}{\st^2} 
  \left\{\frac{1}{\ct^2} B_0(k^2,M_\Hn,M_Z) + 2 
    \sd^2 B_0(k^2,M_\Hpm,M_W) 
  \right.\nonumber\\
  &&
  \left.
    + 2 \frac{(\sd^2-\st^2)^2}{\cd^2} B_0(k^2,M_W,M_W)
  \right\}\nonumber\\
  \Sigma^{\gamma Z}_{(VS)} &=&\frac{\alpha}{4\pi}
  2 M_W^2 \frac{\st^2-\sd^2}{\st\ct} B_0(k^2,M_W,M_W)\nonumber\\
  \Sigma^{WW}_{(VS)} &=& \frac{\alpha}{4\pi} \frac{M_W^2}{\st^2} 
  \left\{ \frac{\sd^2\cd^2}{\ct^2} B_0(k^2,M_Z,M_\Hpm)
    +\frac{(\sd^2-\st^2)^2}{\ct^2}  B_0(k^2,M_Z,M_W)
  \right.\nonumber\\
  &&
  \left.
    + \st^2 B_0(k^2,0,M_W)
    + \cd^2 B_0(k^2,M_\Hn,M_W)
    +  4 \sd^2 B_0(k^2,M_\Kn,M_W)
  \right\}\\
  \mbox{}&&\mbox{}\hspace{13cm}\mbox{}\nonumber
\end{eqnarray} \\

\newpage
\noindent
{\bf Fermion self energies and wave function renormalization:}

\smallskip \noindent
The full list of individual 
Higgs contributions to the fermion self energies
\hfill\\[2ex] 
\centerline{\psfig{figure=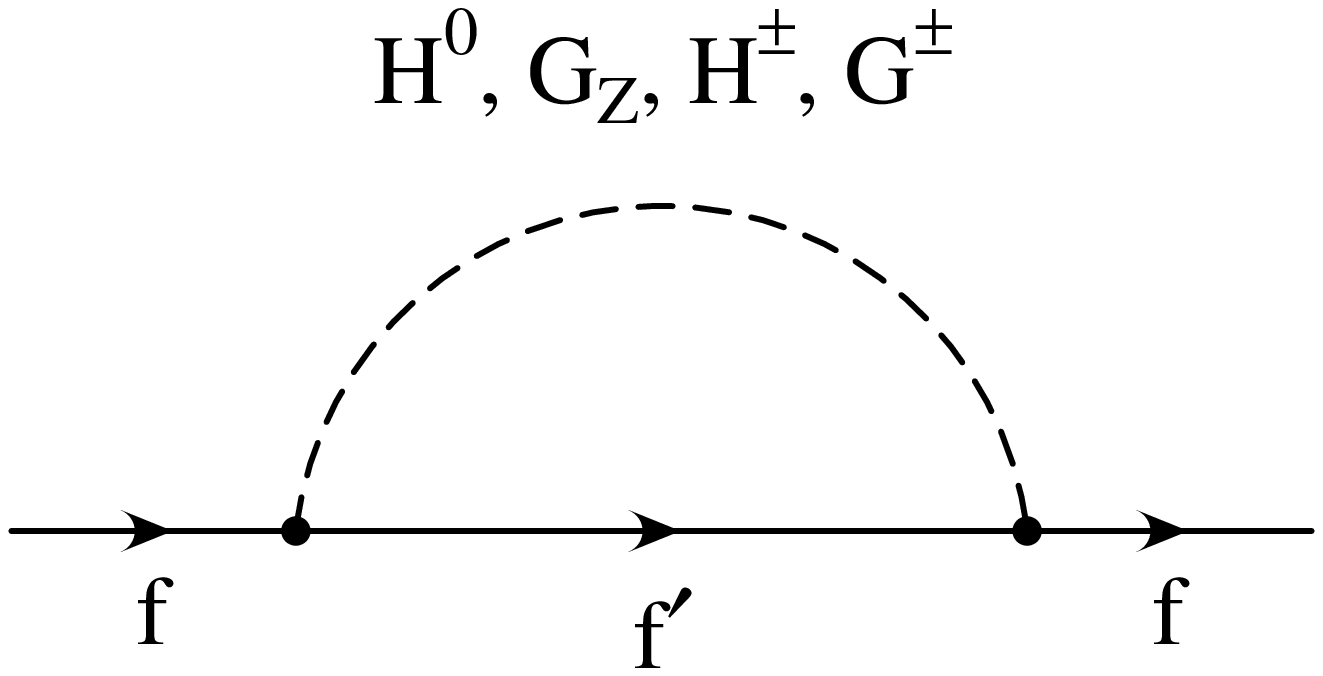,width=5cm}}
\begin{eqnarray}
  \Sigma^f_{(H^0)} &=& - \frac{\alpha}{4\pi} \frac{1}{4 s_\theta^2}
  \frac{m_f^2}{c_\delta^2 M_W^2} 
  \left\{ B_1(k^2,m_f,M_{H^0}) \kslash - B_0(k^2,m_f,M_{H^0}) m_f
  \right\}\nonumber\\
  \Sigma^f_{(G_Z)} &=& - \frac{\alpha}{4\pi} \frac{1}{4 s_\theta^2}
  \frac{m_f^2}{c_\delta^2 M_W^2} 
  \left\{ B_1(k^2,m_f,M_Z) \kslash + B_0(k^2,m_f,M_Z) m_f
  \right\}\nonumber\\
  \Sigma^f_{(G^\pm)} &=& - \frac{\alpha}{4\pi} \frac{1}{4 s_\theta^2}
  \frac{1}{M_W^2} 
  \left\{(m_f^2+m_{\tilde{f}}^2) B_1(k^2,m_{\tilde{f}},M_W) \kslash
  \right.\nonumber\\
  && +(m_f^2-m_{\tilde{f}}^2) B_1(k^2,m_{\tilde{f}},M_W) \kslash \gamma_5
  \nonumber\\
  &&
  \left. + 2 m_{\tilde{f}}^2 B_0(k^2,m_{\tilde{f}},M_W) m_f
  \right\} \nonumber\\
  \Sigma^f_{(H^\pm)} &=& - \frac{\alpha}{4\pi} \frac{1}{4 s_\theta^2}
  \frac{1}{M_W^2} \frac{s_\delta^2}{c_\delta^2} 
  \left\{(m_f^2+m_{\tilde{f}}^2) B_1(k^2,m_{\tilde{f}},M_{H^\pm}) \kslash
  \right.\nonumber\\
  && +(m_f^2-m_{\tilde{f}}^2) B_1(k^2,m_{\tilde{f}},M_{H^\pm}) \kslash
  \gamma_5 \nonumber\\
  &&
  \left. + 2 m_{\tilde{f}}^2 B_0(k^2,m_{\tilde{f}},M_{H^\pm}) m_f
  \right\}\\
  \mbox{}&&\mbox{}\hspace{13cm}\mbox{}\nonumber
\end{eqnarray}
is given here for completeness. In practice, only the charged 
contributions have to be kept for the case $f=b$ because of the internal top 
quark. The neutral contributions are negligibly small also for $f=b$, 
due to the small Yukawa couplings. 
Together with the standard gauge boson contributions, the scalar 
loop diagrams  sum up to the
self energy $\Sigma^f$, decomposed according to
\begin{equation}
 \Sigma^f \ = \ \Sigma^f_V(k^2)\ \kslash \ + \Sigma^f_A(k^2) \
              \kslash \ \gamma_5 \ + m_f \ \Sigma^f_S(k^2)
\label{fself}
\end{equation} 
with scalar functions $\Sigma^f_{V,A,S}$.
The fermion wave function renormalization constants appearing in
eq.~(\ref{vertexct}) read in terms of these functions:
\begin{eqnarray}
  \delta Z_V^f & = & - \ \Sigma^f_V(m_f^2) \ - \  2 m_f^2 \ (
  \Sigma^f_V + \Sigma^f_S)'(m_f^2) \nonumber \\
  \delta Z_A^f & = & \Sigma^f_A(m_f^2) \nonumber \\
  \delta Z_L & = & \delta Z_V + \delta Z_A\quad .   
  \label{zfermion}
\end{eqnarray} \\

\newpage
\noindent
{\bf Vertex corrections:} 

\smallskip \noindent
Neglecting terms proportional to the small
external fermion masses, the 1-loop corrections to the $Z
\bar f f$-vertex contain only vector and axial vector
($\Lambda_{V,A}$)
or left- and right-handed ($\Lambda_{\pm}$) form factors:
\begin{eqnarray}
  \Lambda_\mu^{Zff} &=& \Lambda_+^{Zff}\gamma_\mu\,\frac{1 +
  \gamma_5}{2} +
  \Lambda_-^{Zff}\gamma_\mu\,\frac{1 - \gamma_5}{2} \nonumber\\ 
  & =  & \gamma_\mu
  \left(\Lambda_V^{Zff} -  \gamma_5 \Lambda_A^{Zff}
  \right)\quad.
\label{zff-zerl-einf}
\end{eqnarray}
The form factors $\Lambda_\pm^{Zff}$ consist of the sum of the
contributions given in eqs.~(\ref{vertex1}) -- (\ref{vertex4}) 
with the couplings and 
masses in the attached tables, together with the non-listed
pure gauge boson loops, which are the standard ones.
The entries in the tables contain the
couplings of the fermions to the $Z$ and the Higgs bosons, denoted by
\begin{eqnarray}
  g_f^+ = v_f-a_f &\quad;\quad& g_f^- = v_f+a_f\nonumber\\ 
  g_N^f=-\frac{1}{2 s_\theta}\; \frac{m_f}{M_W c_\delta} &\quad;\quad&
  g_C^f=\frac{1}{\sqrt{2} s_\theta}\; \frac{m_f}{M_W c_\delta}
  \label{v_f+a_f} \quad .
\end{eqnarray}
The arrangements for the couplings, the external momenta  and the
internal  masses are illustrated in the following
figure:\hfill\\  
\centerline{\psfig{figure=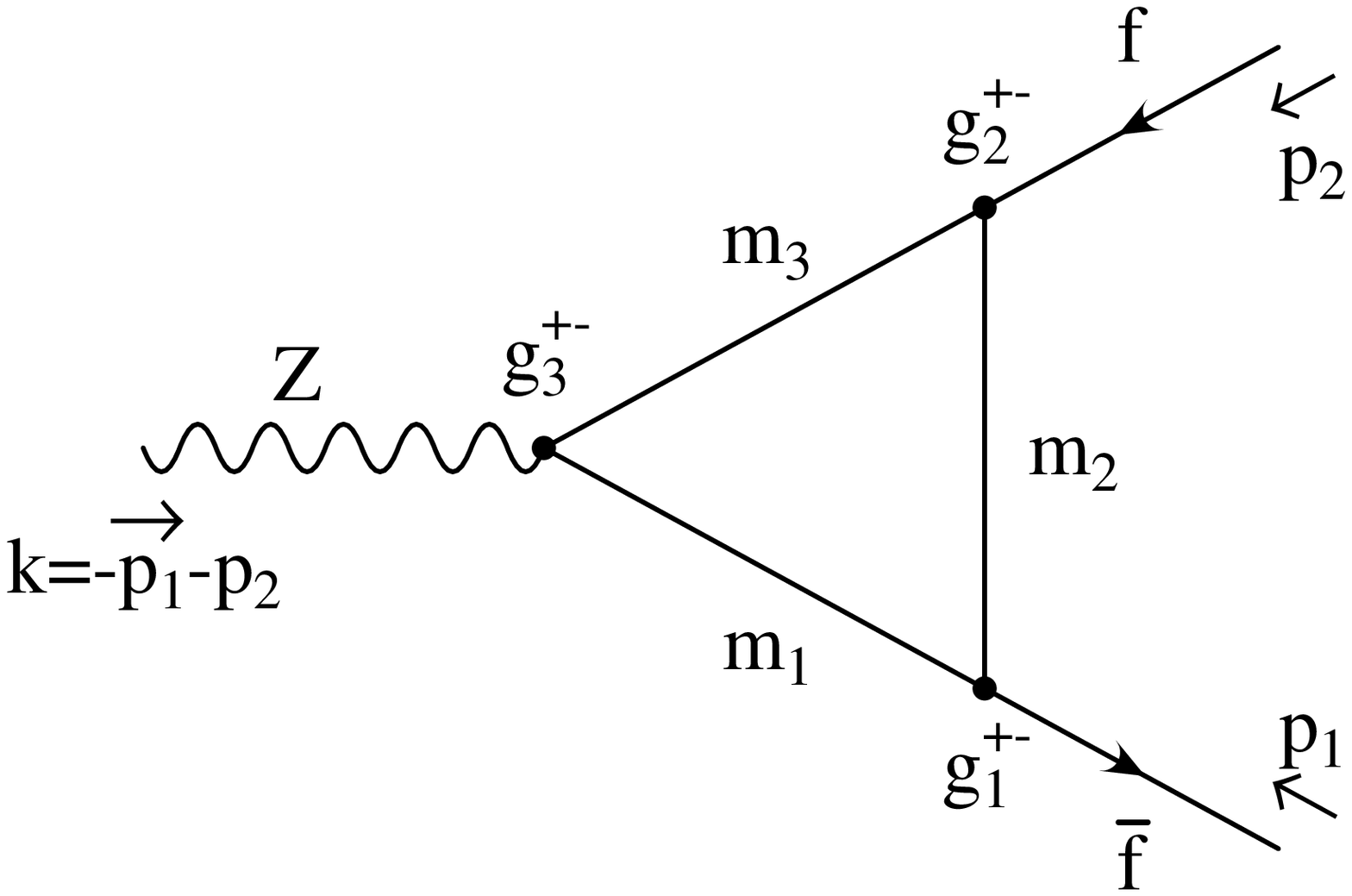,height=5cm}\quad\quad
  \raisebox{2.6cm}{$C = C(p_1,p_2,m_1,m_2,m_3)\quad.$}}
With these conventions,
the individual vertex contributions to the form factors, corresponding
to 4 different topologies, 
read as follows 
[again, as for the fermion self energies, only the contributions with
charged scalars are non-negligible for $b\bar{b}$ finals states;
the others are listed for completeness]:
\hfill\\
\newpage
\centerline{\psfig{figure=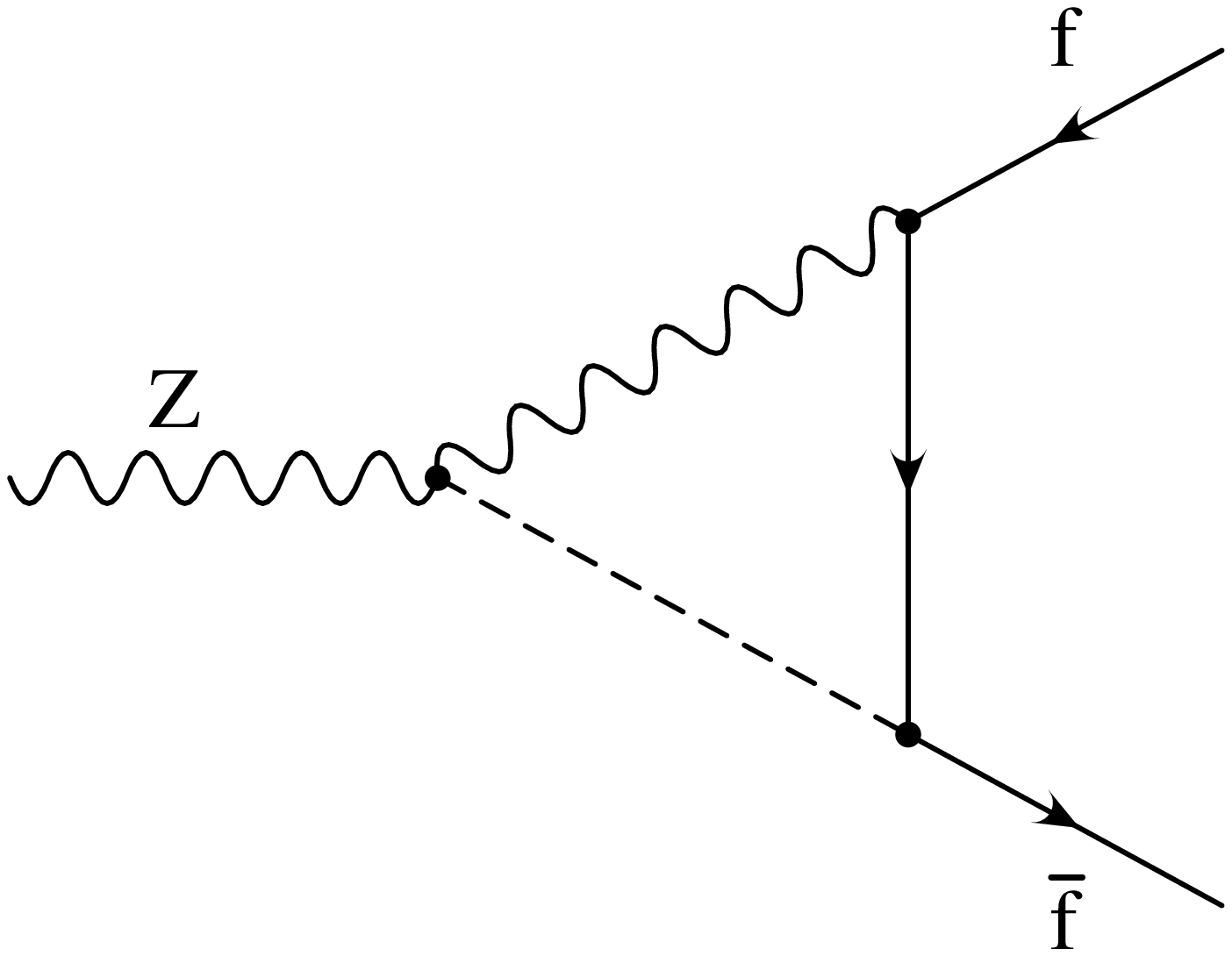,width=5cm}}
\begin{eqnarray}
  \label{vertex1}
  \Lambda_\pm^{Zff} &=& \frac{\alpha}{4\pi} 2 \st \ct \cdot\left\{
  \left[m_f C_{12}
  \right]g_1^\mp g_2^\mp g_3^{\mbox{}}+
  \left[m_f (C_{11} + C_0)
  \right]g_1^\pm g_2^\pm g_3^{\mbox{}}+
  \left[m_{f^\prime} C_0
  \right] g_1^\mp g_2^\pm g_3^{\mbox{}}\right\}\\
  \mbox{}&&\mbox{}\hspace{14cm}\mbox{}\nonumber
\end{eqnarray}
\[
\begin{array}{|c@{}c@{}c||c|c|c|c|c|}
  \hline
  m_1 &\D  m_2 &\D  m_3 &\D  g_1^+ &\D  g_1^- &\D  g_2^+ &\D 
  g_2^- &\D  \quad g_3^+ = g_3^- = g_3^{\mbox{}}\quad \\
  \hline\hline
  \rm H^0 &\D \rm f &\D \rm Z &\D g_N^f &\D g_N^f &\D
  \quad\frac{g_f^+}{2\st\ct}\quad  &\D \quad\frac{g_f^-}{2\st\ct}\quad
  &\D \frac{M_W}{\st\ct^2} \cd \\
  \hline
  \rm H^\pm &\D \rm \tilde{f} &\D \rm W &\D -(2 I_3^{\tilde{f}})
  g_C^{\tilde{f}} \sd &\D - (2 I_3^f) g_C^f \sd &\D 0 &\D
  \frac{1}{\sqrt{2}\st} &\D \frac{M_W}{\st\ct} \sd\cd \\ 
  \hline
  \rm G^\pm &\D \rm \tilde{f} &\D W &\D (2I_3^{\tilde{f}})
  g_C^{\tilde{f}} \cd &\D (2 I_3^f) g_C^f \cd &\D 0 &\D
  \frac{1}{\sqrt{2}\st} &\D\frac{M_W}{\st\ct} (\sd^2-\st^2)\\
  \hline
\end{array}
\]
\centerline{\psfig{figure=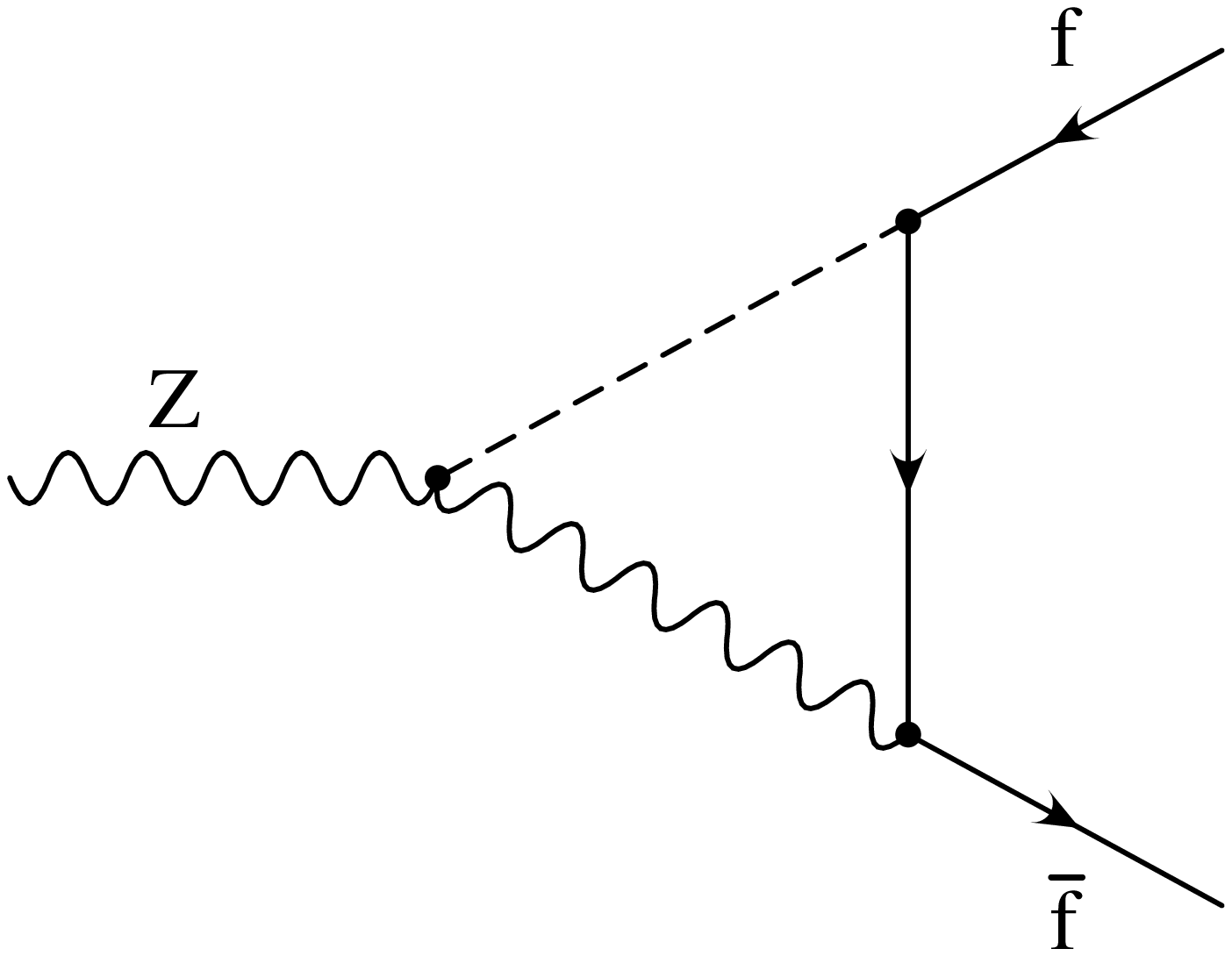,width=5cm}}
\begin{eqnarray}
  \Lambda_\pm^{Zff} &=& \frac{\alpha}{4\pi} 2 \st \ct \cdot\left\{
  \left[-m_f C_{12}
  \right]g_1^\pm g_2^\mp g_3^{\mbox{}}+
  \left[-m_f (C_0 + C_{11})
  \right] g_1^\mp g_2^\pm g_3^{\mbox{}}\right.\nonumber\\
  && +
  \left.\left[m_{f^\prime} C_0
  \right] g_1^\pm g_2^\pm g_3^{\mbox{}}\right\}\\
  \mbox{}&&\mbox{}\hspace{14cm}\mbox{}\nonumber
\end{eqnarray}
\[
\begin{array}{|c@{}c@{}c||c|c|c|c|c|c|}
  \hline
  m_1 &\D  m_2 &\D  m_3 &\D  g_1^+ &\D  g_1^- &\D  g_2^+ &\D 
  g_2^- &\D  \quad g_3^+ = g_3^- = g_3^{\mbox{}} \quad\\
  \hline\hline
  \rm Z &\D \rm f &\D \rm H^0 &\D \quad \frac{g_f^+}{2\st\ct}\quad &\D
  \quad \frac{g_f^-}{2\st\ct} \quad &\D g_N^f &\D g_N^f &\D
  \frac{M_W}{\st\ct^2} \cd \\
  \hline
  \rm W &\D\rm \tilde{f} &\D \rm H^\pm &\D 0 &\D \frac{1}{\sqrt{2}\st}
  &\D - (2 I_3^f) g_C^f \sd &\D -(2 I_3^{\tilde{f}}) g_C^{\tilde{f}} \sd
  &\D \frac{M_W}{\st\ct} \sd\cd \\
  \hline
  \rm W &\D \rm \tilde{f} &\D \rm G^\pm &\D 0 &\D
  \frac{1}{\sqrt{2}\st} &\D (2 I_3^f) g_C^f \cd &\D (2 I_3^{\tilde{f}})
  g_C^{\tilde{f}} \cd &\D \frac{M_W}{\st\ct} (\sd^2-\st^2) \\
  \hline
\end{array}
\]

\centerline{\psfig{figure=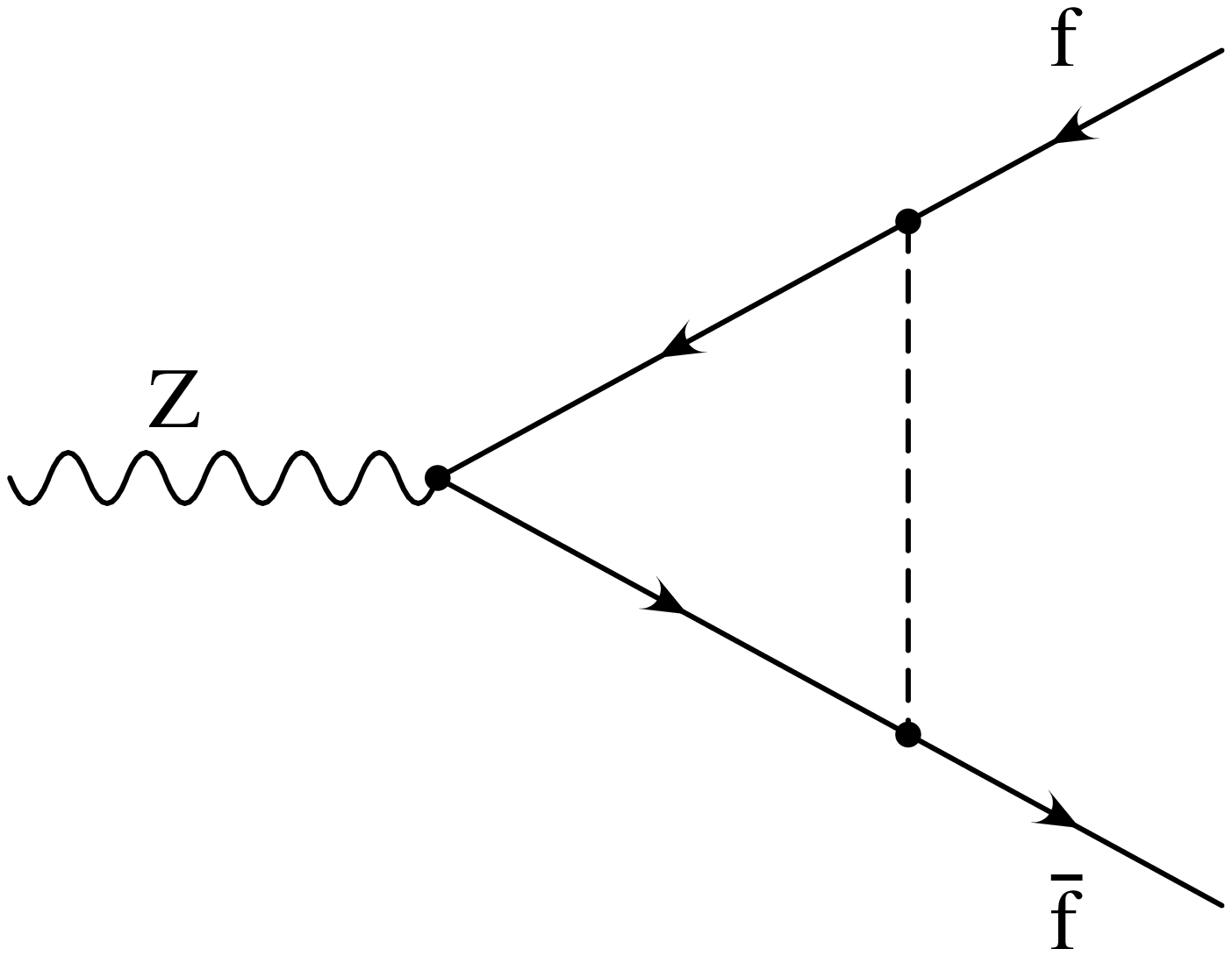,width=5cm}}
\begin{eqnarray}
  \Lambda_\pm^{Zff} &=& \frac{\alpha}{4\pi} 2 \st \ct \cdot\left\{
  \left[(2 C_{20} - \frac{1}{2} ) + m_f^2( C_{11} - C_{12} + C_{21} +
    C_{22} - 2 C_{23})
  \right.\right.\nonumber\\
  && +
  \left.\left. s (C_{12}+C_{23})
  \right] g_1^\mp g_2^\pm g_3^\mp -
  \left[m_{f^\prime}^2 C_0
  \right]g_1^\mp g_2^\pm g_3^\pm-
  \left[m_f^2(C_{12}-C_{11})
  \right]g_1^\pm g_2^\mp g_3^\pm\right.\nonumber\\
  && -
  \left.\left[m_f m_{f^\prime} (C_0+C_{12})
  \right] g_1^\mp g_2^\mp g_3^\pm +
  \left[m_f m_{f^\prime} C_{11}
  \right]g_1^\pm g_2^\pm g_3^\pm +
  \left[m_f m_{f^\prime} C_{12}
  \right]g_1^\mp g_2^\mp g_3^\mp\right. \nonumber\\
  && -
  \left.\left[m_f m_{f^\prime} (C_0 + C_{11})
  \right]g_1^\pm g_2^\pm g_3^\mp\right\}\\
  \mbox{}&&\mbox{}\hspace{14cm}\mbox{}\nonumber
\end{eqnarray}
\[
\begin{array}{|c@{}c@{}c||c|c|c|c|c|c|}
  \hline
  m_1 &\D  m_2 &\D  m_3 &\D  g_1^+ &\D  g_1^- &\D  g_2^+ &\D 
  g_2^- &\D  g_3^+ &\D g_3^- \\
  \hline\hline
  \rm f &\D \rm H^0 &\D \rm f &\D g_N^f &\D g_N^f &\D g_N^f &\D g_N^f &\D
  \quad\frac{g_f^+}{2\st\ct}\quad &\D \quad\frac{g_f^-}{2\st\ct}\quad \\
  \hline
  \rm f &\D \rm G_Z &\D \rm f &\D -i g_N^f (2 I_3^f) &\D i g_N^f (2
  I_3^f) 
  &\D -i g_N^f (2 I_3^f) &\D i g_N^f (2I_3^f) &\D \frac{g_f^+}{2\st\ct}
  &\D \frac{g_f^-}{2\st\ct} \\
  \hline
  \rm \tilde{f} &\D \rm H^\pm &\D \rm \tilde{f} &\D -(2
  I_3^{\tilde{f}}) 
  g_C^{\tilde{f}} \sd &\D -(2 I_3^f) g_C^f \sd &\D -(2 I_3^f) g_C^f \sd
  &\D -(2I_3^{\tilde{f}}) g_C^{\tilde{f}} \sd &\D
  \frac{g_{\tilde{f}}^+}{2\st\ct} &\D \frac{g_{\tilde{f}}^-}{2\st\ct}
  \\
  \hline
  \rm \tilde{f} &\D \rm G^\pm &\D \rm \tilde{f} &\D (2 I_3^{\tilde{f}})
  g_C^{\tilde{f}} \cd &\D (2I_3^f) g_C^f \cd &\D (2 I_3^f) g_C^f \cd
  &\D (2I_3^{\tilde{f}}) g_C^{\tilde{f}} \cd &\D
  \frac{g_{\tilde{f}}^+}{2\st\ct} &\D \frac{g_{\tilde{f}}^-}{2\st\ct}
  \\
  \hline
\end{array}
\]
\newpage
\centerline{\psfig{figure=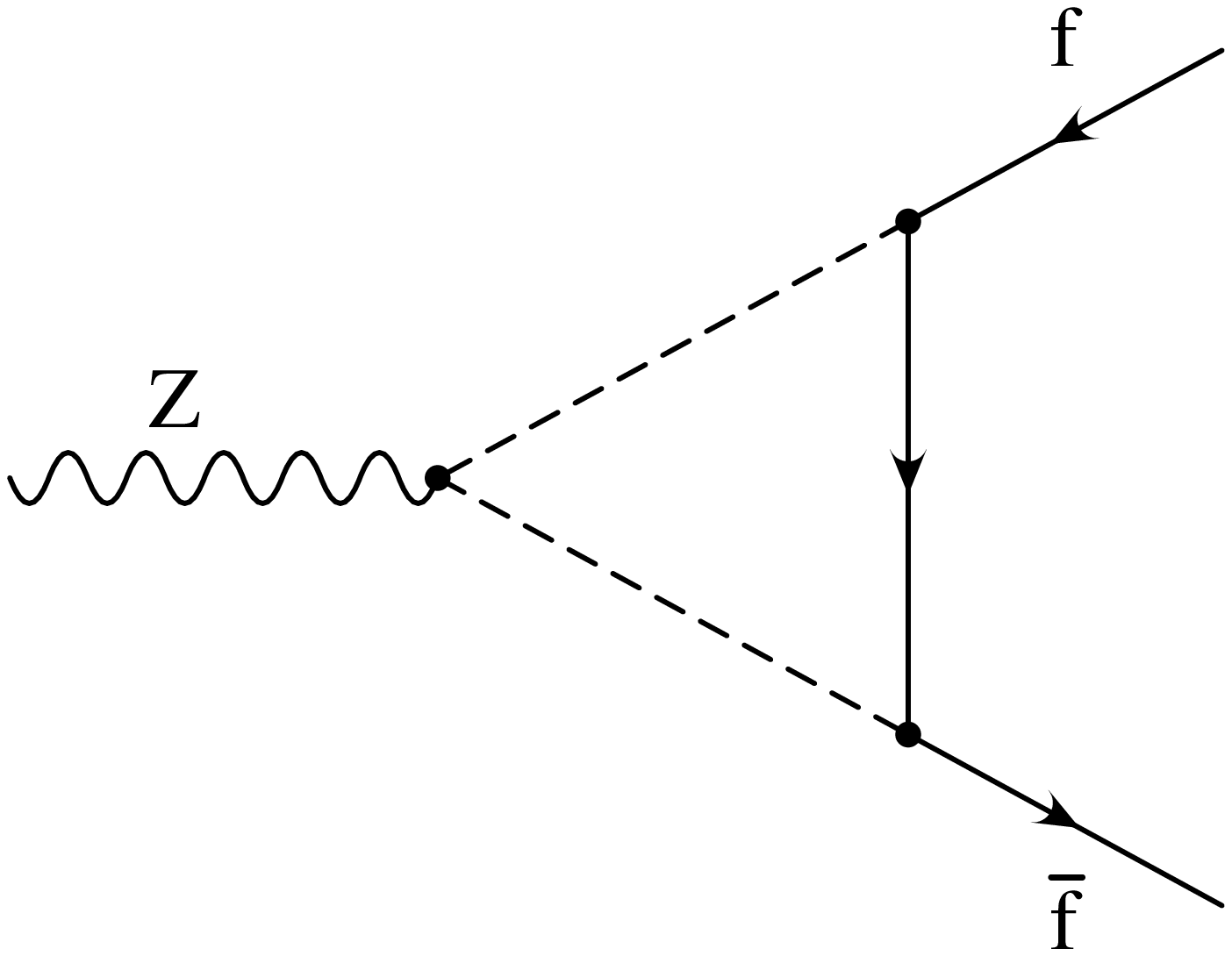,width=5cm}}
\begin{eqnarray}
  \label{vertex4}
  \Lambda_\pm^{Zff} &=& \frac{\alpha}{4\pi} 2 \st \ct \cdot\left\{
  \left[2 C_{20}
  \right]g_1^\mp g_2^\pm g_3^{\mbox{}}\right\}\\
  \mbox{}&&\mbox{}\hspace{14cm}\mbox{}\nonumber
\end{eqnarray}
\[
\begin{array}{|c@{}c@{}c||c|c|c|c|c|}
  \hline
  m_1 &\D  m_2 &\D  m_3 &\D  g_1^+ &\D  g_1^- &\D  g_2^+ &\D 
  g_2^- &\D  g_3^+ = g_3^- = g_3^{\mbox{}}\\
  \hline\hline
  \rm H^0 &\D  \rm f &\D  \rm G_Z &\D  g_N^f &\D  g_N^f &\D  -i g_N^f
  (2I_3^f) &\D 
  i g_N^f (2I_3^f) &\D  -\frac{i}{2\st\ct} \\
  \hline
  \rm G_Z &\D  \rm f &\D  \rm H^0 &\D  -ig_N^f (2I_3^f) &\D  i g_N^f (2
  I_3^f) &\D  g_N^f &\D  g_N^f &\D  \frac{i}{2\st\ct} \\
  \hline
  \rm H^\pm &\D  \rm \tilde{f} &\D  \rm H^\pm &\D  -(2 I_3^{\tilde{f}})
  g_C^{\tilde{f}} s_\delta &\D  -(2I_3^f) g_C^f s_\delta &\D  -(2I_3^f)
  g_C^f s_\delta &\D  -(2 I_3^{\tilde{f}}) g_C^{\tilde{f}} s_\delta &\D 
  (2 I_3^f) \frac{\cd^2-\st^2+\ct^2}{2\st\ct}\\
  \hline
  \rm G^\pm &\D \rm \tilde{f} &\D G^\pm &\D (2I_3^{\tilde{f}})
  g_C^{\tilde{f}} \cd &\D (2I_3^f) g_C^f \cd &\D (2I_3^f) g_C^f \cd &\D
  (2I_3^{\tilde{f}}) g_C^{\tilde{f}} \cd &\D (2I_3^f)\frac{\sd^2 - \st^2 +
    \ct^2}{2\st\ct}\\
  \hline
  \rm H^\pm &\D \rm \tilde{f} &\D \rm G^\pm &\D - (2I_3^{\tilde{f}})
  g_C^{\tilde{f}} \sd &\D - (2I_3^f) g_C^f \sd &\D (2I_3^f)
  g_C^f \cd &\D (2I_3^{\tilde{f}}) g_C^{\tilde{f}} \cd &\D (2I_3^f)
  \frac{\sd\cd}{2\st\ct} \\
  \hline
  \rm G^\pm &\D \rm \tilde{f} &\D \rm H^\pm &\D (2I_3^{\tilde{f}})
  g_C^{\tilde{f}} \cd &\D (2I_3^f) g_C^f \cd &\D - (2I_3^f)
  g_C^f \sd &\D - (2I_3^{\tilde{f}}) g_C^{\tilde{f}} \sd &\D (2I_3^f)
  \frac{\sd\cd}{2\st\ct} \\
  \hline
\end{array}
\]\\[1cm]
In eq. (\ref{vertex1}) to (\ref{vertex4}), 
$f^\prime$ denotes either the fermion $f$ or its isospin
partner $\tilde f$,
dependent on the particle
configuration specified in the attached tables.
\end{appendix}
\end{document}